\documentclass[letterpaper,twocolumn,10pt]{article}
\usepackage{usenix2021}
\usepackage[normalem]{ulem} %
\usepackage{titling}        %
\usepackage{upgreek}        %
\usepackage{url}            %
\usepackage{amssymb}

\usepackage{caption}
\usepackage{etoolbox}
\patchcmd{\thebibliography}
{\settowidth}
{\setlength{\itemsep}{2pt plus 0.1pt}\settowidth}
{}{}

\newtoggle{showmarks}
\toggletrue{showmarks}

\definecolor{myGreen}{rgb}{0,0.7,0}
\iftoggle{showmarks}{
\newcommand\red[1]{\textcolor{red}{#1}}
\newcommand\redstrike[1]{\red{\sout{#1}}}
\newcommand\green[1]{\textcolor{myGreen}{#1}}
\newcommand\greenstrike[1]{\green{\sout{#1}}}
\newcommand\orange[1]{\textcolor{orange}{#1}}
\newcommand\orangestrike[1]{\orange{\sout{#1}}}
\newcommand\blue[1]{\textcolor{blue}{#1}}
\newcommand\bluestrike[1]{\blue{\sout{#1}}}
\newcommand\purple[1]{\textcolor{purple}{#1}}
\newcommand\purplestrike[1]{\purple{\sout{#1}}}
}{
\newcommand\red[1]{#1}
\newcommand\redstrike[1]{\unskip}
\newcommand\green[1]{#1}
\newcommand\greenstrike[1]{\unskip}
\newcommand\orange[1]{#1}
\newcommand\orangestrike[1]{\unskip}
\newcommand\blue[1]{#1}
\newcommand\bluestrike[1]{\unskip}
\newcommand\purple[1]{\unskip}
\newcommand\purplestrike[1]{\unskip}
}

\usepackage{tikz}
\usetikzlibrary{arrows}
\usetikzlibrary{positioning}
\usepackage{authblk}

\makeatletter
\renewcommand\AB@affilsepx{ \protect\Affilfont}
\makeatother
\usepackage{times}
\usepackage{amsmath}
\usepackage{amsthm}
\usepackage{xspace}
\usepackage{microtype}
\usepackage{subcaption}
\usepackage{caption}
\usepackage{xcolor}
\newcommand\todo[1]{\textcolor{red}{TODO: #1}}
\newcommand\edit[1]{#1}

\newcommand{\name}{DispersedLedger\xspace}
\newcommand{\hb}{HoneyBadger\xspace}

\usepackage[subtle, lists=tight, wordspacing=normal, tracking=normal]{savetrees}

\usepackage[small, compact]{titlesec}
\usepackage{titling}
\titlespacing*{\section}{0pt}{4pt}{3pt}
\titlespacing*{\subsection}{0pt}{3pt}{3pt}
\titlespacing*{\subsubsection}{0pt}{3pt}{3pt}

\newtheorem{theorem}{Theorem}[section]
\newtheorem{lemma}[theorem]{Lemma}

\begin{document}

\date{}

\title{\vspace{-2em}\Large\bf
DispersedLedger: High-Throughput Byzantine Consensus \\on Variable Bandwidth Networks}

\author{Lei Yang}
\author{Seo Jin Park}
\author{Mohammad Alizadeh}
\affil{\textit{MIT CSAIL}}
\author{Sreeram Kannan}
\affil{\textit{University of Washington}}
\author{David Tse}
\affil{\textit{Stanford University}}

\maketitle

\begin{abstract}

The success of blockchains has sparked interest in large-scale deployments of Byzantine fault tolerant (BFT) consensus protocols over wide area networks. A central feature of such networks is variable  communication bandwidth across nodes and across time. We present DispersedLedger, an asynchronous BFT protocol that provides near-optimal throughput in the presence of such variable network bandwidth. The core idea of DispersedLedger is to enable nodes to \edit{propose, order, and agree} on blocks of transactions without having to download their full content. By enabling nodes to agree on an ordered log of blocks, with a guarantee that each block is available within the network and unmalleable, \edit{DispersedLedger decouples bandwidth-intensive block downloads at different nodes, allowing each to make progress at its own pace.} We build a full system prototype and evaluate it on real-world and emulated networks. Our results on a geo-distributed wide-area deployment across the Internet shows that DispersedLedger achieves $2\times$ better throughput and 74\% reduction in latency compared to HoneyBadger, the state-of-the-art asynchronous  protocol. 
\end{abstract}

\section{Introduction}

State machine replication (SMR) is a foundational task for building fault-tolerant distributed systems~\cite{lamport-byzantine}. SMR enables a set of nodes to agree on and execute a replicated log of commands (or {\em transactions}). With the success of cryptocurrencies and blockchains, {\em Byzantine fault-tolerant SMR (BFT)} protocols, which tolerate arbitrary behavior from adversarial nodes, have attracted considerable interest in recent years~\cite{hbbft, hotstuff, beat, hyperledger-fabric, casper, tendermint}. The deployment environment for these protocols differs greatly from standard SMR use cases. BFT implementations in blockchain applications must operate over wide-area networks (WAN), among possibly hundreds to thousands of nodes~\cite{hyperledger-fabric,algorand,hbbft}.

Large-scale WAN environments present new challenges for BFT protocols compared to traditional SMR deployments across a few nodes in datacenter. In particular, WANs are subject to {\em variability} in network bandwidth, both across different nodes and across time. 
While BFT protocols are secure in the presence of network variability, their performance can suffer greatly.

To understand the problem, let us consider the high-level structure of existing BFT protocols. BFT protocols operate in epochs, consisting of two distinct phases: (i) a {\em broadcast} phase, in which one or all of the nodes (depending on whether the protocol is leader-based~\cite{hotstuff, synchotstuff} or leaderless~\cite{hbbft, aleph}) broadcast a \textit{block} \edit{(batch of transactions)} to the others; (ii) an {\em agreement} phase, in which the nodes vote for blocks to append to the log, reaching a verifiable agreement (e.g., in the form of a quorum certificate~\cite{pbft}). From a communication standpoint, the broadcast phase is bandwidth-intensive while the agreement phase typically comprises of multiple rounds of short messages that do not require much bandwidth but are latency-sensitive. 

Bandwidth variability hurts the performance of BFT protocols due to {\em stragglers}. In each epoch, the protocol cannot proceed until a super-majority of nodes have downloaded the blocks and voted in the agreement phase. Specifically, a BFT protocol on $N=3f+1$ nodes (tolerant to $f$ faults) requires votes from at least $2f+1$ nodes to make progress~\cite{pbft}. Therefore, the throughput of the protocol is gated by the $(f+1)^{th}$ slowest node in each epoch. The implication is that low-bandwidth nodes (which take a long time to download blocks) hold up the high-bandwidth nodes, preventing them from utilizing their bandwidth efficiently. Stragglers plague even asynchronous BFT protocols~\cite{hbbft}, which aim to track actual network performance (without making timing assumptions), but still require a super-majority to download and vote for blocks in each epoch. We show that this lowers the throughput of these protocols well below the average capacity of the network on real WANs. %

In this paper, we present \name, a new approach to build BFT protocols that significantly improves performance in the presence of bandwidth variability. The key idea behind this approach is to decompose consensus into two steps, one of which is not bandwidth intensive and the other is. First, nodes agree on an ordered log of {\em commitments}, where each commitment is a small digest of a block (e.g., a Merkle root~\cite{merkle}). This step requires significantly less bandwidth than downloading full blocks. Later, each node downloads the blocks in the agreed-upon order and executes the transactions to update its state machine. The principal advantage of this approach is that each node can download blocks at its own pace. Importantly, slow nodes do not impede the progress of fast nodes as long as they have a minimal amount of bandwidth needed to participate in the first step. 

\begin{figure}
\centering
\includegraphics[width=\columnwidth]{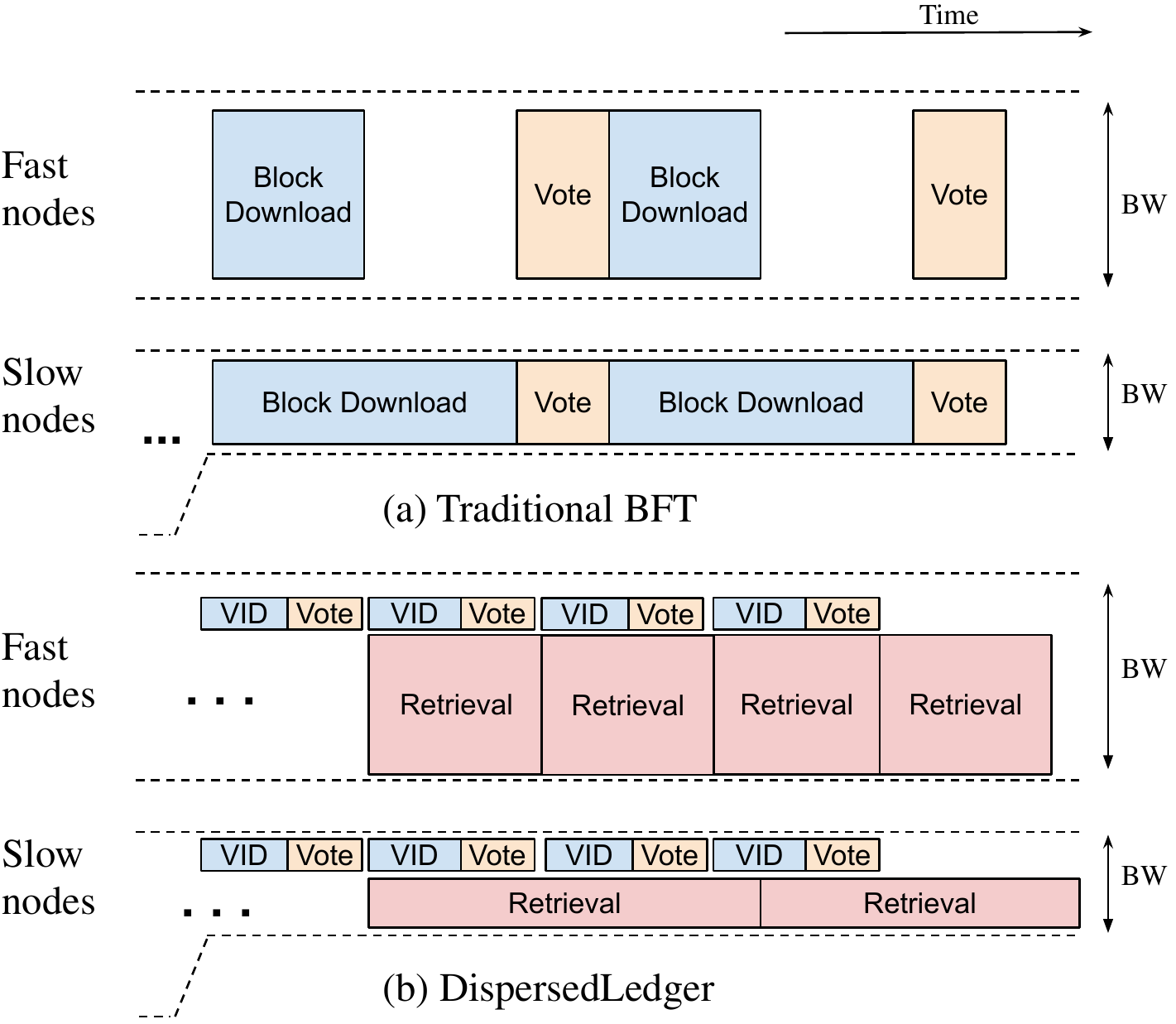}
\caption{Impact of bandwidth variability on overall performance. Bcast: broadcast, Agmt: agreement. Fast nodes currently have a high bandwidth, while slow nodes currently have low bandwidth. (a) In traditional BFT protocols, the speed of consensus is always limited by the slow nodes since they take a long time to download the blocks. (b) \name allows each node to download blocks at its own pace as permitted by its bandwidth.}
\label{fig:idea}
\end{figure}
\edit{The key to realizing this idea is to guarantee the {\em data availability} of blocks}. When a node accepts a commitment into the log, it must know that the block referred to by this commitment {\em is} available in the network and can be downloaded at a later time by any node in the network. Otherwise, an attacker can put a commitment of an unavailable block into the log, thus \edit{halting the system}. To \edit{solve this problem}, our proposal relies on {\em Verifiable Information Dispersal (VID)}~\cite{avid}. VID uses erasure codes to store data across $N$ nodes, such that it can be retrieved later despite Byzantine behavior. Prior BFT protocols like HoneyBadger~\cite{hbbft} have \edit{used VID as a communication-efficient broadcast mechanism}~\cite{avid}, but we use it to guarantee data availability. Specifically, unlike HoneyBadger, nodes in \name do not wait to \edit{download} blocks to vote for them. They vote as soon as they observe that a block has been {\em dispersed}, and the next epoch can begin immediately once there is agreement that dispersal has completed. This enables slow nodes to participate in the latest epoch, even if they fall behind on \edit{block downloads (retrieval)}. Such nodes can catch up on retrievals when their bandwidth improves.  Figure~\ref{fig:idea} shows the structure of \name, contrasting it to traditional BFT protocols.

Enabling nodes to participate in a consensus protocol with minimal bandwidth has applications beyond improving performance on temporally fluctuating bandwidth links. It also creates the possibility of a network with two types of nodes: high-bandwidth nodes and low-bandwidth nodes.  All nodes participate in agreeing on the ordered log of commitments, but only the high-bandwidth nodes retrieve all blocks. Network participants can choose what mode to use at any time. For example, a node running on a mobile device can operate in the low-bandwidth mode when connected to a cellular network, and switch to high-bandwidth mode on WiFi to catch up on block retrievals.  All nodes, both high-bandwidth and low-bandwidth, contribute to the network's security. Our approach is also a natural way to shard a blockchain~\cite{elastico}, where different nodes only retrieve blocks in their own shard.

We make the following contributions:
\begin{itemize}
    \item We propose a new asynchronous VID protocol, AVID-M (\S\ref{sec:avidm-title}). Compared to the current state-of-the-art, AVID-M achieves 1--2 orders of magnitudes better communication cost when operating on small blocks (hundreds of KBs to several MBs) and clusters of more than a few servers.
    \item We design \name (\S\ref{sec:dl-title}), an asynchronous BFT protocol based on \hb \cite{hbbft} with two major improvements: \edit{(i) It decomposes consensus into data availability agreement and block retrieval, allowing nodes to download blocks asynchronously and fully utilize their bandwidth (\S\ref{sec:design-singleEpoch}). (ii) It provides a new solution to the censorship problem~\cite{hbbft} that has existed in such BFT protocols since \cite{acsconstruction} (\S\ref{sec:inter-node-linking}). Unlike \hb, where up to $f$ correct blocks can get dropped every epoch, our solution guarantees that {\em every} correct block is delivered (and executed). The technique is applicable to similarly-constructed protocols, and can improve throughput and achieve censorship resilience without advanced cryptography~\cite{hbbft}.}
    
    \item \edit{We address several practical concerns (\S\ref{sec:discussions}): (i) how to prevent block retrieval traffic from slowing down dispersal traffic, which could reduce system throughput; (ii) how to prevent constantly-slow nodes from falling arbitrarily behind the rest of the network; (iii) how to avoid invalid ``spam'' transactions, now that nodes may not always have the up-to-date system state to filter them out.} 
    
    \item We implement \name in 8,000 lines of Go (\S\ref{sec:impl}) and evaluate it in multiple settings (\S\ref{sec:eval}), including two global testbeds on AWS and Vultr, and controlled network emulations. \name achieves a throughput of 36 MB/s when running at 16 cities across the world, and a latency of 800 ms that is stable across a wide range of load. Compared to \hb, \name has 105\% higher throughput and 74\% lower latency.
\end{itemize}

\section{Background and Related Work}

\subsection{The BFT Problem}

\label{sec:design-model}

\name solves the problem of Byzantine-fault-tolerant state machine replication (BFT)~\cite{lamport-byzantine}. In general, BFT assumes a server-client model, where $N$ servers maintain $N$ replicas of a state machine. At most $f$ servers are Byzantine and may behave arbitrarily. Clients may submit transactions to a correct server to update or read the state machine. 
A BFT protocol must ensure that the state machine is replicated across all correct servers despite the existence of Byzantine servers. 
Usually, this is achieved by delivering a consistent, total-ordered log of transactions to all servers (nodes)~\cite{hbbft}.  
Formally, a BFT protocol provides the following properties:
\begin{itemize}
    \item \textbf{Agreement}: If a correct server executes a transaction $m$, then all correct servers eventually execute $m$.
    \item \textbf{Total Order}: If correct servers $p$ and $q$ both execute transactions $m_1$ and $m_2$, then $p$ executes $m_1$ before $m_2$ if and only if $q$ executes $m_1$ before $m_2$.
    \item \textbf{Validity}: If a correct client submits a transaction $m$ to a correct server, then all correct servers eventually execute $m$.\footnote{Some recent BFT protocols provide a weaker version of validity, which guarantees execution of a transaction $m$ only after being sent to \textit{all} correct servers. This is referred to by different names: ``censorship resilience'' in HoneyBadger, and ``fairness'' in \cite{cachin2001secure, cachin2002secure}.} 
\end{itemize}

There are multiple trust models between BFT servers and the clients.
In this paper, we assume a model used for \textit{consortium blockchains}~\cite{defconsortium, consortium, hyperledger-fabric, goodsurvey}, \edit{where servers and clients belong to organizations. Clients send their transactions through the servers hosted by their organization and trust these servers. %
Many emerging applications of BFT like supply chain tracing \cite{min2019blockchain}, medical data management \cite{azaria2016medrec}, and cross-border transaction clearance \cite{kabra2020mudrachain} fall into this model.}

\subsection{Verifiable Information Dispersal}
\label{sec:bg-avid}

\edit{\name relies on verifiable information dispersal (VID). 
VID resembles a distributed storage, where clients can \textit{disperse} blocks (data files) across servers such that they are available for later \textit{retrieval}. We provide a formal definition of VID in \S\ref{sec:avid-model}.}
The problem of \textit{information dispersal} was first proposed in \cite{rabinid}, where an erasure code was applied to efficiently store a block across $N$ servers without duplicating it $N$ times.%
\cite{firstasyncid} extended the idea to the BFT setting under the asynchrony network assumption. However, it did not consider Byzantine clients; these are malicious clients which try to cause two retrievals to return different blocks. \textit{Verifiable} information dispersal (VID) was first proposed in \cite{avid}, and solved this inconsistency problem. \edit{However, \cite{avid} requires that \textit{every} node downloads the \textit{full} block during dispersal, so it is no more efficient than broadcasting.} The solution was later improved by AVID-FP \cite{avidfp}, which requires each node to only download an $O(1/N)$ fraction of the dispersed data by utilizing fingerprinted cross-checksums \cite{avidfp}. However, because every message in AVID-FP is accompanied by the cross-checksum, the protocol provides low communication cost only when the dispersed data block is much larger than the cross-checksum (about $37N$ bytes). This makes AVID-FP unsuitable for small data blocks and clusters of more than a few nodes. %
In \S\ref{sec:avidm-title}, we revisit this problem and propose AVID-M, a new asynchronous VID protocol that greatly reduces the per-message overhead: from $37N$ bytes to the size of a single hash ($32$ bytes), independent of the cluster size $N$, making the protocol efficient for small blocks and large clusters.%

\subsection{Asynchronous BFT protocols}
\label{sec:async}

\edit{A distributed algorithm has to make certain assumptions on the network it runs on. \name makes the weakest assumption: \textit{asynchrony}~\cite{lynchda}, where messages can be arbitrarily delayed but not dropped. A famous impossibility result \cite{fml} shows there cannot exist a deterministic BFT protocol under this assumption. With randomization, protocols can tolerate up to $f$ Byzantine servers out of a total of $3f+1$ \cite{asyncab1}. \name achieves this bound.}

\edit{Until recently~\cite{hbbft}, asynchronous BFT protocols have been costly for clusters of even moderate sizes because they have a communication cost of at least $O(N^2)$ \cite{cachin2001secure}.} \hb \cite{hbbft} is the first asynchronous BFT protocol to achieve $O(N)$ communication cost per bit of committed transaction (assuming batching of transactions). The main structure of \hb is inspired by \cite{acsconstruction}, and it in turn inspires the design of other protocols including BEAT \cite{beat} and Aleph \cite{aleph}. \edit{In these protocols, all $N$ nodes broadcast their proposed blocks in each epoch, which triggers $N$ parallel Binary Byzantine Agreement (BA) instances to agree on a subset of blocks to commit.  %
\cite{avid} showed that VID can be used as an efficient construction of reliable broadcast, by invoking retrieval immediately after dispersal. \hb and subsequent protocols use this construction as a blackbox. BEAT \cite{beat} explores multiple tradeoffs in \hb and proposes a series of protocols based on the same structure. One protocol, BEAT3, also includes a VID subcomponent. However, BEAT3 is designed to achieve BFT \textit{storage}, which resembles a distributed key-value store.}

\subsection{Security Model}
\label{sec:model}
\edit{
Before proceeding, we summarize our security model. We make the following assumptions:
\begin{itemize}
    \item The network is asynchronous (\S\ref{sec:async}).
    \item The system consists of a fixed set of $N$ nodes (servers). A subset of at most $f$ nodes are Byzantine, and $N \ge 3f+1$. $N$ and $f$ are protocol parameters, and are public knowledge.
    \item Messages are authenticated using public key cryptography. The public keys are public knowledge.
\end{itemize}
}

\section{AVID-M: An Efficient VID Protocol}
\label{sec:avidm-title}
\subsection{Problem Statement}

\label{sec:avid-model}

\edit{
VID provides the following two primitives:
$\mathtt{Disperse}(B)$, which a client invokes to disperse block $B$, and
$\mathtt{Retrieve}$, which a client invokes to retrieve 
block $B$. 
Clients invoke the $\mathtt{Disperse}$ and $\mathtt{Retrieve}$ primitives against a particular {\em instance} of VID, where each VID instance is in charge of dispersing a different block.
Multiple instances of VID may run concurrently and independently. To distinguish between these instances, clients and servers tag all messages of each VID instance with a unique ID for that instance. For each instance of VID, each server triggers a $\mathtt{Complete}$ event to indicate that the dispersal has completed.}

\edit{
A VID protocol must provide the following properties~\cite{avid} for each instance of
VID:
\begin{itemize}
    \item \textbf{Termination}: If a correct client invokes $\mathtt{Disperse}(B)$ and no other client invokes $\mathtt{Disperse}$ on the same instance, then all correct servers eventually $\mathtt{Complete}$ the dispersal. 
    \item \textbf{Agreement}: If some correct server has $\mathtt{Complete}$d the dispersal, then all correct servers eventually $\mathtt{Complete}$ the dispersal.
    \item \textbf{Availability}: If a correct server has $\mathtt{Complete}$d the dispersal, and a correct client invokes $\mathtt{Retrieve}$, it eventually reconstructs some block $B'$.
    \item \textbf{Correctness}: If a correct server has $\mathtt{Complete}$d the dispersal, then correct clients always reconstruct the \textit{same} block $B'$ by invoking $\mathtt{Retrieve}$. Also, if a correct client initiated the dispersal by invoking $\mathtt{Disperse}(B)$ and no other client invokes $\mathtt{Disperse}$ on the same instance, then $B=B'$.
\end{itemize}
}

\subsection{Overview of AVID-M}

At a high level, a VID protocol works by encoding the dispersed block using an erasure code and storing the encoded \textit{chunks} across the servers. A server knows a dispersal has completed when it hears from  enough peers that they have received their chunks. To retrieve a dispersed block, a client can query the servers to obtain the chunks and decode the block. Here, one key problem is verifying the correctness of encoding. Without verification, a malicious client may distribute \textit{inconsistent} chunks that \edit{have more than one decoding result} depending on which subset of chunks are used for decoding, violating the Correctness property. As mentioned in \S\ref{sec:bg-avid}, AVID \cite{avid} and AVID-FP solve this problem by requiring servers to download the chunks or fingerprints of the chunks from all correct peers and examine them during dispersal. While this eliminates the possibility of inconsistent encoding, the extra data download required limits the scalability of these protocols.

\begin{figure}
\centering
\includegraphics[width=\columnwidth]{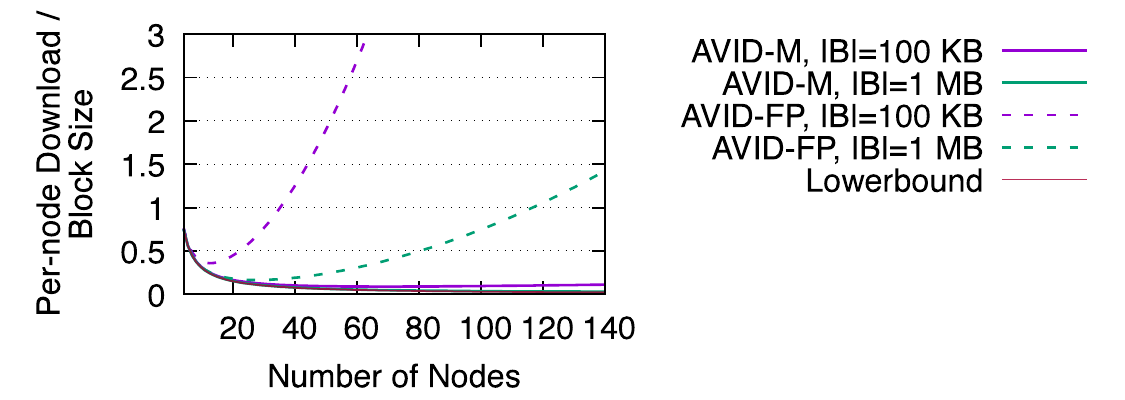}
\caption{ \textit{Per-node} communication cost during dispersal of AVID-M and AVID-FP normalized over the size of the dispersed block. At $N=128$ (the biggest cluster size in our evaluation), every node in AVID-M downloads as much as $1/32$ of a block, while a node in AVID-FP downloads $1.2\times$ the size of the full block.}
\label{fig:vid-comparison}
\end{figure}

\edit{More specifically, while AVID-FP~\cite{avidfp} can achieve optimal communication complexity as the block size $|B|$ goes to infinity, its overhead for practical values of $|B|$ and $N$ (number of servers) can be quite high.} This is because every message in AVID-FP is accompanied by a fingerprinted cross-checksum~\cite{avidfp}, which is $N\lambda + (N-2f)\gamma$ in size. Here, $\lambda, \gamma$ are security parameters, and we use $\lambda=32$ bytes, $\gamma=16$ bytes as suggested by \cite{avidfp}.  The key factor that limits the scalability of AVID-FP is that the size of the cross-checksum is proportional to $N$. Combined with the fact that a node receives $O(N)$ messages during dispersal, the overhead caused by cross-checksum increases quadratically as $N$ increases. Fig.~\ref{fig:vid-comparison} shows the impact of this overhead. At $N > 40$, $|B|=100$ KB, every node needs to download more than the \textit{full size} of the block being dispersed.

\edit{
We develop a new VID protocol for the asynchronous network model, Asynchronous Verifiable Information Dispersal with Merkle-tree (AVID-M). AVID-M is based on one key observation: as long as clients can independently verify the encoding during {\em retrieval}, the servers do not need to do the verification during dispersal.  In AVID-M, a client invoking $\mathtt{Disperse}(B)$ commits to the set of (possibly inconsistent) chunks using a short, constant-sized commitment $H$. Then the server-side protocol simply agrees on $H$ and guarantees enough chunks that match $H$ are stored by correct servers. This can be done by transmitting only $H$ in the messages, compared to the $O(N)$-sized cross-checksums in AVID-FP. During retrieval, a client verifies that the block it decodes produces the same commitment $H$ when re-encoded.
}

Since AVID-M's per-message overhead is a small constant (32 bytes), it can scale to many nodes without requiring a large block size. \edit{In fact, AVID-M achieves a per-node communication cost of $O(|B|/N+\lambda N)$, much lower than AVID-FP's $O(|B|/N+\lambda N^2 + \gamma N^2)$.}  Fig.~\ref{fig:vid-comparison} compares AVID-M with AVID-FP. At $|B|=1$ MB, AVID-M is close to the theoretical lowerbound\footnote{Each node has to download at least $\frac{1}{N-2f}$-fraction of the dispersed data. This is to prevent a specific attack: a malicious client sends chunks to all $f$ malicious servers plus $N-2f$ honest servers. For now the malicious servers do not deviate from the protocol, so the protocol must terminate (otherwise it loses liveness). Then the malicious servers do not release the chunks, so the original data must be constructed from the $N-2f$ chunks held by honest servers, so each honest server must receive an $\frac{1}{N-2f}$-fraction share.} even at $N > 100$, while AVID-FP stops to provide any bandwidth saving (compared to every server downloading full blocks) after $N>120$. \edit{Finally, we note that both AVID-M and AVID-FP rely on the security of the hash. So with the same hash size $\lambda$, AVID-M is no less secure than AVID-FP.}  

\subsection{AVID-M Protocol}
\label{sec:avidm-protocol}

The \textbf{Dispersal algorithm} is formally defined in  Fig.~\ref{fig:dispersalAlgo}.
A client initiates a dispersal by encoding the block $B$ using an $(N-2f, N)$-erasure code and constructing a Merkle tree~\cite{merkle} out of the encoded chunks. The root $r$ of the Merkle tree is a secure summary of the array of the chunks. The client sends one chunk to each server along with the Merkle root $r$ and a Merkle \textit{proof} that proves the chunk belongs to root $r$.
\edit{Servers then need to make
sure at least $N-2f$ chunks under the same Merkle root are stored at \textit{correct} servers for retrieval. To do that, servers exchange a round of $\mathtt{GotChunk}(r)$ messages to announce
the reception of the chunk under root $r$.
When $N-f$ servers have announced $\mathtt{GotChunk}(r)$, they know at least $N-2f$ correct
servers have got the chunk under the same root $r$, so they exchange a round of
$\mathtt{Ready}(r)$ messages to collectively $\mathtt{Complete}$ the dispersal.}

The \textbf{Retrieval algorithm} is formally defined in Fig~\ref{fig:retrievalAlgo}.
A client begins retrieval by requesting chunks for the block from all servers. Servers respond by providing the chunk, the Merkle root $r$, and the Merkle proof proving that the chunk belongs to the tree with root $r$. Upon collecting $N-2f$ different chunks with the same root, the client can decode and obtain a block $B'$. \edit{However, the client must ensure that other retrieving clients
also obtain $B'$ no matter which subset of $N-2f$ chunks they use -- letting
clients perform this check is a key idea of AVID-M. To do that,
the client re-encodes $B'$, constructs a Merkle tree out of the resulting chunks, and verifies that the root is the same as $r$.
If not, the client returns an error string as the retrieved content.}

\begin{figure}
\centering
\fbox{%
  \parbox{0.95\columnwidth}{%

\underline{\textbf{$\mathtt{Disperse}(B)$ invoker}}

\begin{enumerate}
	\item Encode the input block $B$ using an $(N-2f, N)$-erasure code, which results in N chunks, $C_1, C_2, \dots, C_N$.
    \item Form a Merkle tree with all chunks $C_1, C_2, \dots, C_N$, and calculate the Merkle tree root, $r$.
    \item Send $\mathtt{Chunk}(r, C_i, P_i)$ to the $i$-th server. Here $P_i$ is the Merkle proof showing $C_i$ is the $i$-th chunk under root $r$.
\end{enumerate}

\underline{\textbf{$\mathtt{Disperse}(B)$ handler for the $i$-th server}}
\begin{itemize}
    \item Upon receiving $\mathtt{Chunk}(r, C_i, P_i)$ from a client:
    \begin{enumerate}
        \item Check if $C_i$ is the $i$-th chunk under root $r$ by verifying the proof $P_i$. If not, ignore the message.
	\item Set $\mathtt{MyChunk} = C_i$, $\mathtt{MyProof} = P_i$, $\mathtt{MyRoot} = r$ (all initially unset).
	\item Broadcast $\mathtt{GotChunk}(r)$ if it has not sent a $\mathtt{GotChunk}$ message before.
    \end{enumerate}
    \item Upon receiving $\mathtt{GotChunk}(r)$ from the $j$-th server:
    \begin{enumerate}
        \item Increment $\mathtt{ShareCount}[r]$ (initially $0$).
	\item If $\mathtt{ShareCount}[r] \ge N-f$, broadcast $\mathtt{Ready}(r)$. %
    \end{enumerate}
    \item Upon receiving $\mathtt{Ready}(r)$ from the $j$-th server:
    \begin{enumerate}
        \item Increment $\mathtt{ReadyCount}[r]$ (inititally $0$).
        \item If $\mathtt{ReadyCount}[r] \ge f+1$, broadcast $\mathtt{Ready}(r)$. %
	\item If $\mathtt{ReadyCount}[r] \ge 2f+1$, set $\mathtt{ChunkRoot}=r$. Dispersal is $\mathtt{Complete}$.
    \end{enumerate}
\end{itemize}
} }
\caption{
	\edit{
		Algorithm for $\mathtt{Disperse}(B)$. Servers ignore duplicate messages (same sender and same type). When broadcasting, servers also send the message to themselves.}
}
\label{fig:dispersalAlgo}
\end{figure}

\begin{figure}
\centering
\fbox{%
  \parbox{0.95\columnwidth}{%

\underline{\textbf{$\mathtt{Retrieve}$ invoker}}

\begin{itemize}
    \item Broadcast $\mathtt{RequestChunk}$ to all servers.
    \item Upon getting $\mathtt{ReturnChunk}(r, C_i, P_i)$ from the $i$-th server:
    \begin{enumerate}
        \item Check if $C_i$ is the $i$-th chunk under root $r$ by verifying the proof $P_i$. If not, ignore the messsage.
        \item Store the chunk $C_i$ with the root $r$.
    \end{enumerate}
    \item Upon collecting $N-2f$ or more chunks with the same root $r$:
    \begin{enumerate}
	    \item Decode using any $N-2f$ chunks with root $r$ to get a block $B'$. Set $\mathtt{ChunkRoot}=r$ (initially unset).
	    \item Encode the block $B'$ using the same erasure code to get chunks ${C_1}', {C_2}', \dots, {C_N}'$.
	    \item Compute the Merkle root $r'$ of $C_1', C_2', \dots, C_N'$.
	    \item Check if $r' = \mathtt{ChunkRoot}$. If so, return $B'$. Otherwise, return string ``BAD\_UPLOADER''.
    \end{enumerate}
\end{itemize}

\underline{\textbf{$\mathtt{Retrieve}$ handler for the $i$-th server}}
\begin{itemize}
	\item Upon receiving $\mathtt{RequestChunk}$,  respond with message $\mathtt{ReturnChunk}(\mathtt{ChunkRoot}, \mathtt{MyChunk}, \mathtt{MyProof})$ if $\mathtt{MyRoot} = \mathtt{ChunkRoot}$. Defer responding if dispersal is not $\mathtt{Complete}$ or any variable here is unset.
\end{itemize}

} }
	\caption{\edit{
		Algorithm for $\mathtt{Retrieve}$. Clients ignore duplicate messages (same sender and same type).}
}
\label{fig:retrievalAlgo}
\end{figure}

\edit{The AVID-M protocol described in this section provides the four properties mentioned in \S\ref{sec:avid-model}.
We provide a proof sketch for each property, and point to Appendix \ref{sec:avidm-correctness}
for complete proofs.}

\noindent\edit{
\textbf{Termination (Theorem \ref{thm:avid-termination}).}
A correct client sends correctly encoded chunks to all servers with root $r$.
The $N-f$ correct servers will broadcast $\mathtt{GotChunk}(r)$ upon getting
their chunk. All correct servers will receive the $N-f$ $\mathtt{GotChunk}(r)$  and send out
$\mathtt{Ready}(r)$, so all correct servers will receive at least $N-f$ $\mathtt{Ready}(r)$. 
Because $N-f > 2f + 1$, all correct servers will $\mathtt{Complete}$.}

\noindent\edit{\textbf{Agreement (Theorem \ref{thm:avid-agreement}).}
A server $\mathtt{Complete}$s after receiving $2f+1$ $\mathtt{Ready}(r)$,
of which $f+1$ must come from correct servers. So all correct servers will
receive at least $f+1$ $\mathtt{Ready}(r)$. This will drive all of them
to send $\mathtt{Ready}(r)$.
Eventually every correct server will receive
$N-f$ $\mathtt{Ready}(r)$, which is enough to $\mathtt{Complete}$ ($N - f > 2f+1$).} 

\noindent\edit{\textbf{Availability (Theorem \ref{thm:avid-availability}).}
To retrieve, a client must collect $N-2f$ chunks with the {\em same} root. This requires that at least $N-2f$ correct servers have a chunk for the same root. Now suppose that a correct server $\mathtt{Complete}$s when receiving $2f+1$ $\mathtt{Ready}(r)$.
When this happens, at least one correct server has sent $\mathtt{Ready}(r)$. We prove that this implies that at least $N-2f$ correct servers must have sent $\mathtt{GotChunk}(r)$ (Lemma \ref{thm:avid-ready}),i.e., they have received the chunk. Assume the contrary. Then there will be
less than $N-f$ $\mathtt{GotChunk}(r)$. Now a correct server only sends $\mathtt{Ready}(r)$ if it either receives (i) at least $N-f$ $\mathtt{GotChunk}(r)$, or (ii) at least $f+1$ $\mathtt{Ready}(r)$. Neither is possible (see Lemma~\ref{thm:avid-ready}).} 

\edit{All correct servers agree on the same root upon $\mathtt{Complete}$ by setting $\mathtt{ChunkRoot}$ to the same value (Lemma~\ref{thm:avid-agree-root}). To see why, notice that each server will only send one $\mathtt{GotChunk}$ per instance. If correct servers $\mathtt{Complete}$ with 2 (or more) $\mathtt{ChunkRoot}$s, then at least $N-f$ servers must have sent $\mathtt{GotChunk}$ for each of these roots. But $2(N-f) > N+f$, hence at least one correct server must have sent $\mathtt{GotChunk}$ for two different roots, which is not possible.}

\noindent\edit{\textbf{Correctness (Theorem \ref{thm:avid-correctness}).}
First, note that two correct clients finishing $\mathtt{Retrieve}$ will set
$\mathtt{ChunkRoot}$ to be the same, i.e., they will decode from chunks under the same Merkle
root $r$ (Lemma \ref{thm:avid-agree-root}). However, we don't know if two different subsets of
chunks under $r$ would decode to the same block, because a malicious client could
disperse arbitrary data as chunks. To ensure consistency of $\mathtt{Retrieve}$ across
different correct clients, every correct client re-encodes the decoded block $B'$, calculates
the Merkle root $r'$ of the encoding result, and compares $r'$ with the root $r$.
There are two possibilities: (i) Some correct client gets $r' = r$. Then $r$ corresponds to the chunks given by the {\em correct} encoding of $B'$, so every correct client decoding from any subset of blocks under $r$ will also get $B'$ and $r' = r$. (ii) No correct client gets $r' = r$, i.e, all of them get $r' \ne r$. In this case, they all deliver the fixed error string. In either case, all correct clients return the same data (Lemma \ref{thm:avid-root-summary}).
}

\section{DispersedLedger Design}
\label{sec:dl-title}

\subsection{Overview}
\label{sec:ba-def}

\begin{figure}
\centering
\includegraphics[width=\columnwidth]{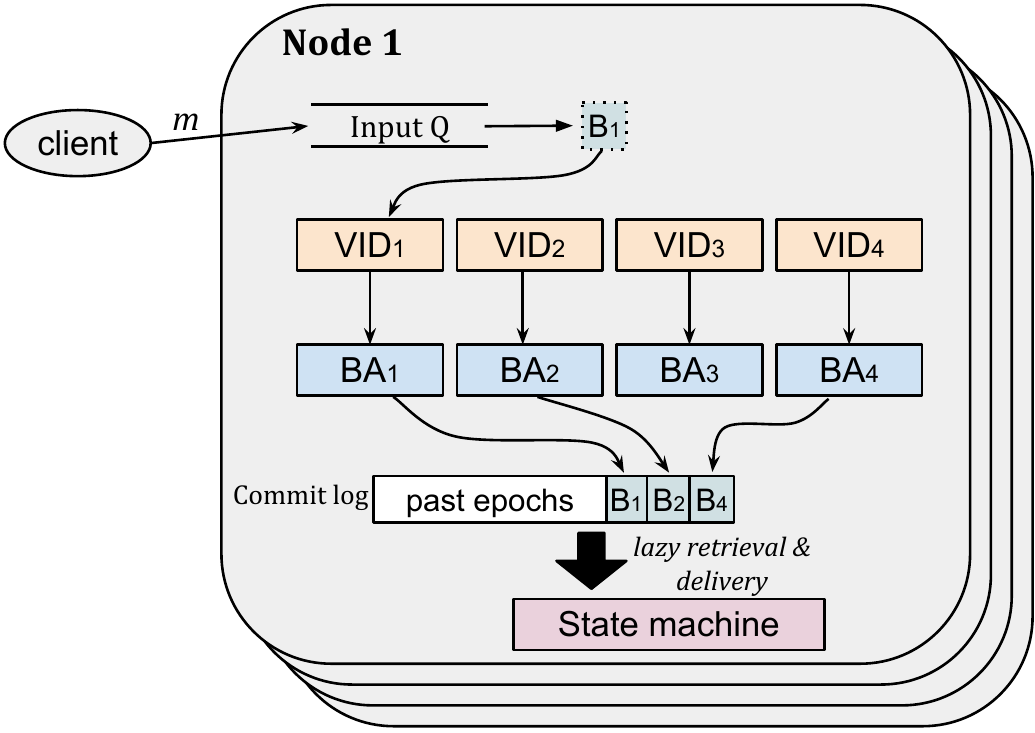}
	\caption{\name architecture with $N=4$. During this single epoch, $4$ VIDs are initiated, one for each node, and three blocks $B_1$,$B_2$ and $B_4$ are \edit{committed}. }
\label{fig:architecture}

\end{figure}

\name is a modification of HoneyBadger \cite{hbbft}, a state-of-the-art asynchronous BFT protocol. 
\edit{\hb runs in epochs, where each epoch commits between $N-f$ to $N$ blocks (at most 1 block from each node).
As shown in Fig.~\ref{fig:architecture}, transactions submitted by clients are stored in each node's input queue. At the beginning of each epoch, every node creates a block from transactions in its input queue, and proposes it to be committed to the log in the current epoch. Once committed, all transactions in the block will eventually be retrieved and delivered to the state machine for execution.}

\edit{\name has two key differences with \hb. First, unlike \hb, a node in \name does not broadcast its proposed block; instead, it \textit{disperses} the proposed block among the entire cluster using \edit{AVID-M (which we will refer to as VID from here on)}. As shown in Fig.~\ref{fig:architecture}, there are $N$ instances of VID in every epoch, one for each node. \name then relies on $N$ instances of Binary Agreement (BA, details below)~\cite{baprotocol} to reach a consensus on which proposed blocks have been successfully dispersed and thus should be committed in the current epoch. \edit{Once committed, a block can be retrieved by nodes lazily at any time (concurrently with future block proposals and dispersals).} The asynchronous retrieval of blocks allows each node to adapt to temporal network bandwidth variations by adjusting the rate it retrieves blocks without slowing down other nodes.}

\edit{In \hb, up to $f$ correct blocks can be dropped in every epoch (\S\ref{sec:inter-node-linking}). This wastes bandwidth and can lead to censorship where blocks from certain nodes are always dropped~\cite{hbbft}. \name's second innovation is a new method, called inter-node linking, that guarantees every correct block is committed.}

\name uses an existing BA protocol~\cite{baprotocol} that completes in $O(1)$ time (parallel rounds) with $O(N\lambda)$ per-node communication cost, where $\lambda$ is the security parameter. \edit{In BA, each node provides a binary  $\mathtt{Input}(\{0, 1\})$ as input to the protocol, and may get an $\mathtt{Output}(\{0, 1\})$ event indicating the result of the BA instance. %
Formally, a BA protocol has the following properties}:
\begin{itemize}
    \item \textbf{Termination}: If all correct nodes invoke $\mathtt{Input}$, then every correct node eventually gets an $\mathtt{Output}$.
	\item \textbf{Agreement}: If any correct node gets $\mathtt{Output}(b)$ ($b \in \{0, 1\}$), then every correct node eventually gets $\mathtt{Output}(b)$.
	\item \textbf{Validity}: If any correct node gets  $\mathtt{Output}(b)$ ($b \in \{0, 1\}$), then at least one correct node has invoked $\mathtt{Input}(b)$.
\end{itemize}

\subsection{Single Epoch Protocol}
\label{sec:design-singleEpoch}

\edit{In each epoch, the goal is to agree on a set of (the indices of) at least $N-f$ dispersed blocks which are available for later retrieval. An epoch contains $N$ instances of VID and BA. Let $\text{VID}_i^e$ be the $i$-th ($1 \le i \le N$) VID instance of epoch $e$. $\text{VID}_i^e$ is reserved for the $i$-th node to disperse (propose) its block.\footnote{\edit{Correct nodes ignore attempts from another node $j$ ($j \ne i$) to disperse into $\text{VID}_i^e$ by dropping $\mathtt{Chunk}$ messages for $\text{VID}_i^e$ from node $j$ ($j \ne i$). Therefore, a Byzantine node cannot impersonate and disperse blocks on behalf of others.}} Let $\text{BA}_i^e$ be the $i$-th ($1 \le i \le N$) BA instance of epoch $e$. $\text{BA}_i^e$ is for agreeing on whether to commit the block dispersed by the $i$-th node.
}

\begin{figure}
\centering
\fbox{%
  \parbox{0.95\columnwidth}{%
\underline{\textbf{Phase 1. Dispersal} at the $i$-th server}
\begin{enumerate}
    \item Let $B_i^e$ be the block to disperse (propose) for epoch $e$.
    \item Invoke $\mathtt{Disperse}(B_i^e)$ on $\text{VID}_i^e$ (acting as a client).
\end{enumerate}
\begin{itemize}
	\item Upon $\mathtt{Complete}$ of $\text{VID}_j^e$ ($1 \le j \le N$), if we have not invoked $\mathtt{Input}$ on $\text{BA}_j^e$, invoke $\mathtt{Input}(1)$ on  $\text{BA}_j^e$.
	\item Upon $\mathtt{Output}(1)$ of least $N-f$ BA instances, invoke $\mathtt{Input}(0)$ on all remaining BA instances on which we have not invoked $\mathtt{Input}$.
	\item Upon $\mathtt{Output}$ of all BA instances,
		\begin{enumerate}
			\item Let (local variable) $S^e_i \subset \{1 \dots N\}$ be the indices of all BA instances that $\mathtt{Output}(1)$. That is, $j \in S_i^e$ if and only if $\text{BA}_j^e$ has $\mathtt{Output}(1)$ at the $i$-th server.
			\item Move to retrieval phase.
		\end{enumerate}
\end{itemize}

\underline{\textbf{Phase 2. Retrieval}}

\begin{enumerate}
	\item For all $j \in S^e_i$, invoke $\mathtt{Retrieve}$ on $\text{VID}^e_j$ to download full block ${B_j^e}'$.
    \item Deliver $\{{B_j^e}' | j \in S^e_i\}$ (sorted by increasing indices).
\end{enumerate}

}
} 
\caption{
	\edit{Algorithm for single-epoch \name.}
}
\label{fig:singleEpochAlgo}

\end{figure}

\edit{Fig.~\ref{fig:singleEpochAlgo} describes the single epoch protocol for the $i$-th node at epoch $e$. It begins by taking the block $B_i^e$ to be proposed for this epoch, and dispersing it for epoch $e$ through $\text{VID}_i^e$. Note that every block in the system is dispersed using a unique VID instance identified by its epoch number and proposing node.}

\edit{Nodes now need to decide which blocks get committed in this epoch, and they should only commit blocks that have been successfully dispersed. Because there are potentially $f$ Byzantine nodes, we cannot wait for all $N$ instances of VID to complete because Byzantine nodes may never initiate their VID $\mathtt{Disperse}$. On the other hand, nodes cannot simply wait for and commit the first $N-f$ VIDs to $\mathtt{Complete}$, because VID instances may $\mathtt{Complete}$ in different orders at different nodes (hence correct nodes would not be guaranteed to commit the same set of blocks). \name uses a strategy first proposed in \cite{acsconstruction}. Nodes use $\text{BA}_i^e$ to explicitly agree on whether to commit $B_i^e$ (which should be dispersed in $\text{VID}_i^e$). Correct nodes input $1$ into $\text{BA}_i^e$ only when $\text{VID}_i^e$ $\mathtt{Complete}$s, so $\text{BA}_i^e$ outputs $1$ only if $B_i^e$ \textit{is} available for later retrieval. When $N-f$ BA instances have output $1$, nodes give up on waiting for any more VID to $\mathtt{Complete}$, and input $0$ into the remaining BAs to explicitly signal the end of this epoch. This is guaranteed to happen because VID instances of the $N-f$ correct nodes will always $\mathtt{Complete}$ by the Termination property (\S\ref{sec:avid-model}). Once the set of committed blocks are determined, nodes can start retrieving the full blocks. After all blocks have been downloaded, a node sorts them by index number and delivers (executes) them in order. }

The single-epoch \name protocol is readily chained together epoch by epoch to achieve full SMR, as pictured in Fig.~\ref{fig:architecture}. At the beginning of every epoch, a node takes transactions from the head of the input buffer to form a block. After every epoch, a node checks if its block is committed. If not, it puts the transactions in the block back to the input buffer and proposes them in the next epoch. Also, a node delivers epoch $e$ only after it has delivered all previous epochs.

\subsection{Inter-node Linking} 
\label{sec:inter-node-linking}

\edit{\noindent\textbf{Motivation.}} An important limitation of the aforementioned single-epoch protocol (and all protocols with a similar construction \cite{hbbft, beat}) is that not all proposed blocks from correct nodes are committed in an epoch. An epoch only guarantees to commit $N-f$ proposed blocks, out of which $N-2f$ are guaranteed to come from correct nodes. In other words, at most $f$ blocks proposed by correct nodes are dropped every epoch. Dropped blocks can happen \textit{with or without} adversarial behavior. Transmitting such blocks wastes bandwidth, for example, reducing \hb's throughput by 25\% in our experiments (\S\ref{sec:internet-performance}). To make the matter worse, the adversary (if present) can determine which blocks to drop \cite{hbbft}, i.e. at most $f$ correct servers can be \textit{censored} such that \textit{no} block from these servers gets committed. \edit{\hb provides a partial mitigation by keeping the proposed blocks encrypted until they are committed so that the adversary cannot censor blocks by their content. The adversary {\em can}, however, censor blocks based on the proposing node.\footnote{\edit{\hb suggests sending transactions to all nodes to prevent censorship, but this isn't possible for consortium blockchains and still wastes bandwidth due to dropped blocks (\S\ref{sec:internet-performance})}.}  This is unacceptable for consortium blockchains (\S\ref{sec:model}), because the adversary could censor \textit{all} transactions from certain (up to $f$) organizations. Moreover, \hb's mitigation relies on threshold cryptography, which incurs a high computational cost \cite{beat}.}

\begin{figure}
\centering
\includegraphics[width=\columnwidth]{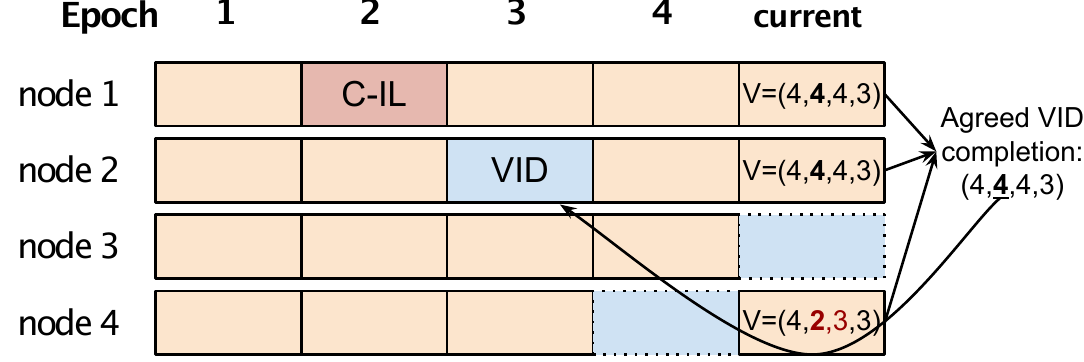}
\caption{An example of commits by inter-node linking where $N=4, f=1$. Each box indicates a block proposed by a node at an epoch. Orange blocks are committed by BA. ``VID'' indicates that the block is dispersed but not committed. ``C-IL'' indicates a block committed by inter-node linking. Blue dotted boxes indicate a VID in progress. In the current epoch, after delivering the blocks from node 1, 2, and 4, the block proposed in epoch 3 by node 2 will be delivered by inter-node linking.}
\label{fig:internodeLinking}

\end{figure}

\noindent\edit{\textbf{Our solution.} We propose a novel solution to this problem, called \textit{inter-node linking}, that guarantees \textit{all} blocks from correct nodes are committed. Inter-node linking eliminates any censorship or bandwidth waste, and is readily applicable to similarly constructed protocols like \hb and BEAT. Notice that a block not committed by BA in a given epoch may still finish its VID. For example, in Fig.~\ref{fig:internodeLinking}, the block proposed by node 2 in epoch 3 was dispersed but did not get selected by BA in that epoch. The core idea is to have nodes identify such blocks and deliver them in a consistent manner in later epochs.}

\edit{Each node $i$ keeps track of which VID instances have $\mathtt{Complete}$d, in the form of an array $V^e_i$ of size $N$, which stores the local view at that node. 
When node $i$ starts epoch $e$, it populates $V^e_i[j]$ (for all $1 \le j \le N$) with the largest epoch number such that all node $j$'s VID instances up to epoch $V^e_i[j]$ have completed. For example, in Fig.~\ref{fig:internodeLinking}, $(4,4,4,3)$ would be a valid array $V$ for the current epoch, and would indicate that node 2's VID for epoch 3 has completed but node 4's VID in epoch 4 has not.}%

\edit{Each node $i$ reports its local array $V^e_i$ in the block $B^e_i$ it proposes in each epoch (in addition to the normal block content). As shown in Fig.~\ref{fig:internodeLinking}, the BA mechanism then commits at least $N-f$ blocks in each epoch. During retrieval for epoch $e$, a node first retrieves the blocks committed by BA in epoch $e$ and delivers (executes) them as in the single-epoch protocol (\S\ref{sec:design-singleEpoch}). It then extracts the set of $V$ arrays in the committed blocks, i.e. $\{V^e_j | j \in S^e_i\}$, and combines the information across these arrays to determine additional blocks that it should retrieve (and deliver) in this epoch. Note that $S^e_i = S^e_j$ for any two correct nodes $i, j$ due to the Agreement property of BA, so all correct nodes will use the same set of observations and get the same result.\footnote{If a particular $\mathtt{Retrieve}$ returns string ``BAD\_UPLOADER'' or the block is ill formatted, we use array $[\infty, \infty, \dots, \infty]$ as the observation.}}

\edit{Using the committed $V$ arrays, the inter-node linking protocol computes an epoch number $E^e[j]$ for each node $j$. This is computed locally by each node $i$, but we omit the index $i$ since all (correct) nodes compute the same value.  Each node then retrieves and delivers (executes) all blocks from node $j$ until epoch $E^e[j]$. To ensure total order, nodes sort the blocks, first by epoch number then by node index. They also keep track of blocks that have been delivered so that no block is delivered twice.}

\edit{In computing $E^e[j]$, we must be careful to not get misled by Byzantine nodes who may report arbitrary data in their $V$ arrays. For example, naively taking the largest value reported for node $j$ across all $V$ arrays, i.e., $\max_{k \in S^e_i} V^e_k[j]$, would allow a Byzantine node to fool others into attempting to retrieve blocks that do not exist. Instead, we take the $(f+1)^\text{th}$-largest value; this guarantees that at least one correct node $i$ has reported in its array $V^e_i$ that node $j$ has completed all its VIDs up to epoch $E^e[j]$. Recall that by the Availability property of VID (\S\ref{sec:avid-model}), this ensures that these blocks are available for retrieval. Also, since all correct blocks eventually finish VID (Termination property), all of them will eventually be included in $E^e$ and get delivered. We provide pseudocode for the full \name protocol in Appendix~\ref{sec:appendix-internodeLinking}.}

\subsection{Correctness of \name}
\label{sec:dl-security}
\edit{
We now analyze the correctness of the \name protocol by showing it guarantees the three properties required for BFT (\S\ref{sec:design-model}). %
Full proof is in Appendix \ref{sec:dlproof}.
}

\noindent\edit{
\textbf{Agreement and Total Order (Theorem \ref{thm:agreement}).} Transactions are embedded in blocks, so we only need
to show Agreement and Total Order of \textit{block} delivery at each correct node.
Blocks may get committed and delivered through two mechanisms: BA and inter-node linking. First consider blocks committed by BA. BA's Agreement and VID's Correctness properties guarantee that (i) all correct nodes will retrieve the same set of blocks for each epoch, and (ii) they will download the same content for each block. Now consider the additional blocks committed by inter-node linking. As discussed in \S\ref{sec:inter-node-linking}, correct nodes determine these blocks based on identical information ($V$ arrays) included in the blocks delivered by BA. Hence they all retrieve and deliver the same set of blocks (Lemma \ref{thm:counter-agreement}). Also, all correct nodes use the same sorting criteria (BA-delivered blocks sorted by node index, followed by inter-node-linked blocks sorted by epoch number and node index), so they deliver blocks in the same order.
}

\noindent\edit{
\textbf{Validity (Theorems \ref{thm:progress}, \ref{thm:validity}).}
Define ``correct transactions'' as ones submitted by correct clients
to correct nodes (servers). We want to prove every correct transaction is eventually delivered (executed).
This involves two parts: (i) correct nodes do not hang, so that every correct transaction
eventually gets proposed in some correct block (Theorem \ref{thm:progress});
(ii) all correct blocks eventually get delivered (Theorem \ref{thm:validity}).
}

\edit{
For part (i), note that all BAs eventually $\mathtt{Output}$, since in every epoch
at least $N-f$ BAs will $\mathtt{Output}(1)$ (Lemma \ref{thm:ba-progress}), and then all correct nodes will $\mathtt{Input}(0)$ to the remaining BAs and drive them to termination. Further, 
all blocks selected by BA or inter-node linking are guaranteed to be successfully dispersed, so $\mathtt{Retrieve}$ for them will eventually finish. By BA's Validity property, a BA only produces $\mathtt{Output}(1)$ when some correct node has $\mathtt{Input}(1)$, which can only happen if that node sees the corresponding VID $\mathtt{Complete}$. Also, as explained in \S\ref{sec:inter-node-linking}, inter-node linking only selects blocks that at least one correct node observes to have finished dispersal (Lemma \ref{thm:correct-estimation}). By the Availability property of VID (\S\ref{sec:avid-model}), all these blocks are available for retrieval.}
\edit{
For part (ii), note that all correct blocks eventually finish VID (Termination property). The inter-node linking protocol will therefore eventually identify all such blocks to have completed dispersal (Lemma \ref{thm:correct-estimation}) and deliver them (if not already delivered by BA).}

\subsection{Practical Considerations}
\label{sec:discussions}
\noindent
\edit{\textbf{Running multiple epochs in parallel.}
In \name, nodes perform dispersal sequentially, proceeding to the dispersal phase for the next epoch as soon as the dispersal for the current epoch has completed (all BA instances have $\mathtt{Output}$). On the other hand, the retrieval phase of each epoch runs asynchronously at all nodes. To prevent slow nodes from stalling the progress of fast nodes, it is important that they participate in dispersal at as high a rate as possible, using only remaining bandwidth for retrieval. This effectively requires prioritizing dispersal traffic over retrieval traffic when there is a network bottleneck. Furthermore, a node can retrieve blocks from multiple epochs in parallel (e.g., to increase network utilization), but it must always deliver (execute) blocks in a serial order. Ideally, we want to fully utilize the network but prioritize traffic for earlier epochs over later epochs to minimize delivery latency.
Mechanisms to enforce prioritization among different types of messages are implementation-specific (\S\ref{sec:impl}).}

\noindent\edit{\textbf{Constantly-slow nodes.}
Since \name decouples the progress of fast and slow nodes, a natural question is:
what if some nodes are constantly slow and do not have a chance to catch up? The possibility of some nodes constantly lagging behind is a common concern for BFT protocols. A BFT protocol cannot afford to wait for the slowest servers, because they could be Byzantine servers trying to stall the system \cite{slownodes}. Therefore the slow servers (specifically the $f$ slowest servers) can be left behind, unable to catch up. Essentially, there is a tension between accommodating servers that are correct but slow, and preventing Byzantine nodes from influencing the system.}

\edit{\name expands this issue beyond the $f$ slowest servers. We discuss two simple mitigations. First, the system designer could mandate a minimum average bandwidth per node such that all correct nodes can support the target system throughput over a certain timescale $T$. Every node must support the required bandwidth over time $T$ but can experience lower bandwidth temporarily without stalling other nodes. Second, correct nodes could simply stop proposing blocks when too far behind, e.g., if their retrieval is more than $P$ epochs behind the current epoch ($P = 1$ is the same as \hb). If enough nodes fall behind and stop proposing, it automatically slows down the system. A designer can choose parameters $T$ or $P$ to navigate the tradeoff between bandwidth variations impacting system throughput and how far behind nodes can get.}

\noindent\edit{\textbf{Spam transactions.} In \name, nodes do not check the validity of blocks they propose, deferring this check to the retrieval phase. This creates the possibility of malicious servers or clients spamming the system with invalid blocks. }

\edit{
Server-sent spam cannot be filtered even in conventional BFT protocols, because by the time other servers download the spam blocks, they have already wasted bandwidth. Similarly, \hb must perform BA (and incur its compute and bandwidth cost) regardless of the validity of the block, because by design, all BAs must eventually finish for the protocol to make progress~\cite{hbbft}. Therefore, server-sent spam harms \name and \hb equally. Fortunately, server-sent spam is bounded by the fraction of Byzantine servers ($f/N$).
}

\edit{
On the other hand, client-sent spam is not a major concern in consortium blockchains (\S\ref{sec:design-model}). In consortium blockchains, the organization is responsible for its clients, and a non-Byzantine organization would not spam the system.\footnote{\edit{A Byzantine organization could of course spam, but this is the same as the server-sent spamming scenario, in which \name is no worse than \hb.}} For these reasons, some BFT protocols targeting consortium blockchains such as HyperLedger Fabric~\cite{hyperledger-fabric} forgo transaction filtering prior to broadcast for efficiency and privacy gains.}

\edit{
In more open settings, where clients are free to contact any server, spamming is a concern. A simple modification to the \name protocol enables the same level of spam filtering as \hb. Correct nodes simply stop proposing new transactions when they are lagging behind in retrieval. Instead, they propose an empty block (with no transactions) to participate in the current epoch. In this way, correct nodes only propose transactions when they can verify them. Empty blocks still incur some overhead, so a natural question is: what is the performance impact of these empty blocks? Our results show that it is minor and this variant of \name, which we call ``DL-Coupled'', retains most of the performance benefits (\S\ref{sec:spam-eval}).}

\section{Implementation}
\label{sec:impl}

We implement \name in 8,000 lines of Go. The core protocol of \name is modelled as 4 nested IO automata: \texttt{BA}, \texttt{VID}, \texttt{DLEpoch}, and \texttt{DL}. \texttt{BA} implements the binary agreement protocol proposed in \cite{baprotocol}. \texttt{VID} implements our verifiable information dispersal protocol AVID-M described in \S\ref{sec:avidm-protocol}. We use a pure-Go implementation of Reed-Solomon code \cite{rslibrary} for encoding and decoding blocks, and an embedded key-value storage library \cite{kvstore} for storing blocks and chunks. \texttt{DLEpoch} nests $N$ instances of \texttt{VID} and \texttt{BA} to implement one epoch of \name (\S\ref{sec:design-singleEpoch}). Finally, \texttt{DL} nests multiple instances of \texttt{DLEpoch} and the inter-node linking logic (\S\ref{sec:inter-node-linking}) to implement the full protocol.

\noindent \textbf{Traffic prioritization.} 
\edit{Prioritizing dispersal traffic over retrieval is made complicated because nodes cannot be certain of the bottleneck capacity for different messages and whether they share a common bottleneck.} For example, rate-limiting the low-priority traffic may result in under-utilization of the network. Similarly, simply enforcing prioritization between each individual pair of nodes may lead to significant priority inversion if two pairs of nodes share the same bottleneck. In our implementation, we use a simple yet effective approach to achieve prioritization in a work conserving manner (without static rate limits) \edit{inspired by MulTcp~\cite{multcp}}. For each pair of nodes, we establish two connections, and we modify the parameters of the congestion control algorithm of one connection so that it behaves like $T$ ($T>1$) connections . We then send high-priority traffic on this connection, and low-priority traffic on the other (unmodified) connection. At all bottlenecks, the less aggressive low-priority connection will back off more often and yield to the more aggressive high-priority connection. On average, a high-priority connection receives $T$ times more bandwidth than a competing low-priority connection at the same bottleneck.\footnote{\edit{Similar approaches have been used in other usecases to control bandwidth sharing among competing flows~\cite{vikram}.}} Note that in \name, high-priority traffic consists of only a tiny fraction of the total traffic that a node handles ($1/20$ to $1/10$ in most cases as shown in \S\ref{sec:eval-scalability}), and its absolute bandwidth is low. Therefore our approach will not cause congestion to other applications competing at the same bottleneck. In our system, we set $T=30$. We use QUIC as the underlying transport protocol and modify the \texttt{quic-go}~\cite{quic-go} library to add the knob $T$ for tuning the congestion control.

To prioritize retrieval traffic by epoch, we order retrieval traffic on a per-connection basis by using separate QUIC streams for different epochs. We modify the scheduler \texttt{quic-go}~\cite{quic-go} to always send the stream with the lowest epoch number.

\noindent \textbf{Rate control for block proposal.} \name requires some degree of batching to amortize the fixed cost of BA and VID. However, if unthrottled, nodes may propose blocks too often and the resulting blocks could be very small, causing low bandwidth efficiency. More importantly, since dispersal traffic is given high priority, the system may use up all the bandwidth proposing inefficient small blocks and leave no bandwidth for block retrieval. To solve this problem, our implementation \edit{employs a simple form of adaptive batching \cite{stout}. Specifically, we} limit the block proposal rate using Nagle's algorithm~\cite{nagle}. A node only proposes a new block if (i) a certain duration has passed since the last block was proposed, or (ii) a certain amount of data has accumulated to be proposed in the next block. In our implementation, we use 100 ms as the delay threshold, and 150 KB as the size threshold. This setup works well for all of our evaluation experiments.

\section{Evaluation}
\label{sec:eval}

Our evaluation answers the following questions:

\begin{enumerate}
    \item What is the throughput and the latency of \name in a realistic deployment?
    \item Is \name able to consistently achieve good throughput regardless of network variations?
    \item How does the system scale to more nodes?
\end{enumerate}

We compare \name (DL) with the original \hb (HB) and our optimized version: \hb-Link. \hb-Link (HB-Link) combines the inter-node linking in \name with \hb, so that every epoch, all (instead of $N-2f$) honest blocks are guaranteed to get into the ledger. \edit{We also experiment with DL-Coupled, a variant of \name where nodes only propose new transactions when they are up-to-date with retrievals  (\S\ref{sec:discussions}).}

\subsection{Experimental Setup}

We run our evaluation on AWS EC2. In our experiments, every node is hosted by an EC2 \texttt{c5d.4xlarge} instance with 16 CPU cores, 16 GB of RAM, 400 GB of NVMe SSD, and a 10 Gbps NIC. The nodes form a fully connected graph, i.e. there is a link between every pair of nodes. We run our experiments on two different scenarios. First, a \textit{geo-distributed} scenario, where we launch VMs at 16 major cities across the globe, one at each city. We don't throttle the network. This scenario resembles the typical deployment of a \edit{consortium} blockchain. In addition, we measure the throughput of the system on another testbed on Vultr (details are in Appendix \ref{apx:vultr-throughput}). Second, a \textit{controlled} scenario, where we start VMs in one datacenter and apply artificial delay and bandwidth throttling at each node using Mahimahi~\cite{mahimahi}. Specifically, we add a one-way propagation delay of 100~ms between each pair of nodes to mimic the typical latency between distant major cities \cite{pingstats}, and model the ingress and egress bandwidth variation of each node as independent Gauss-Markov processes (more details in \S\ref{sec:controlled-experiments}). This controlled setup allows us to precisely define the variation of the network condition and enables fair, reproducible evaluations. Finally, to generate the workload for the system, we start a thread on each node that generates transactions in a Poisson arrival process.

\subsection{Performance over the Internet}
\label{sec:internet-performance}

First, we measure the performance of \name on our geo-distributed testbed and compare it with \hb.

\begin{figure}
\centering
\includegraphics[width=\columnwidth]{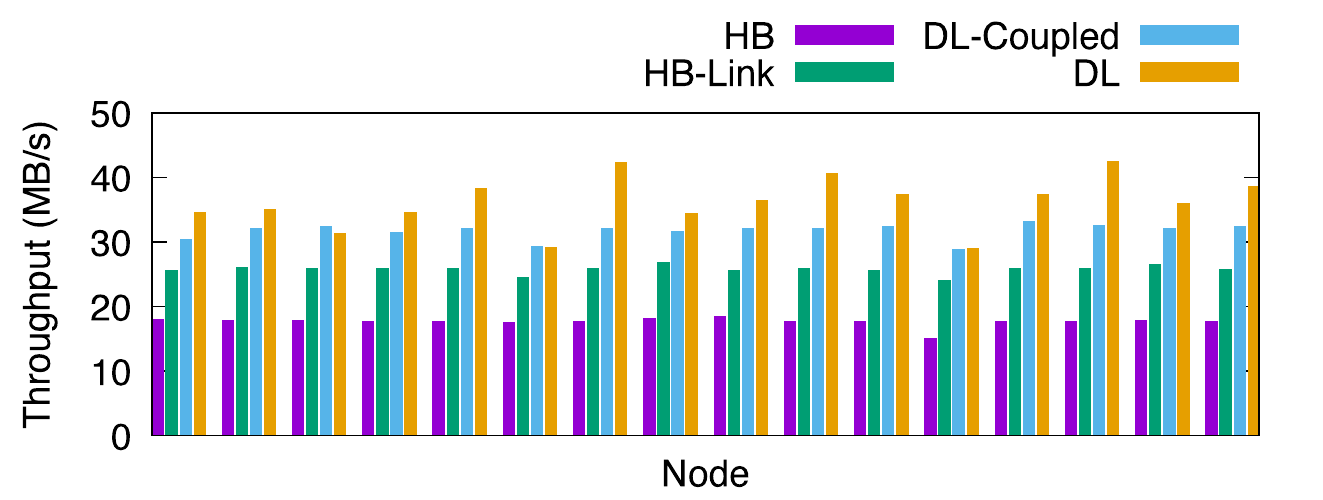}
\caption{\label{fig:geo-thruput} Throughput of each server running different protocols on the geo-distributed setting. 
}

\end{figure}

\noindent \textbf{Throughput.} To measure the throughput, we generate a high load on each node to create an infinitely-backlogged system, and report the rate of confirmed transactions at each node. Because the internet bandwidth varies at different locations, we expect the measured throughput to vary as well.  Fig.~\ref{fig:geo-thruput} shows the results. 
\name achieves on average \edit{105\%} better throughput than \hb. To confirm that our scheme is robust, we also run the experiment on another testbed using a low-cost cloud vendor. Results in \S\ref{apx:vultr-throughput} show that \name significantly improves the throughput in that setting as well.

\name gets its throughput improvement mainly for two reasons. First, inter-node linking ensures all blocks that successfully finish VID get included in the ledger, so no bandwidth is wasted. In comparison, in every epoch of \hb at most $f$ blocks may \textit{not} get included in the final ledger. The bandwidth used to broadcast them is therefore wasted. As a result, inter-node linking provides at most a factor of $N/(N-f)$ improvement in effective throughput. To measure the gain in the real-world setting, we modify \hb to include the same inter-node linking technique and measure its throughput. Results in Fig.~\ref{fig:geo-thruput} show that enabling inter-node linking provides a \edit{45\%} improvement in throughput on our geo-distributed testbed.

\begin{figure}
\centering
\begin{subfigure}[b]{0.5\linewidth}
         \centering
         \includegraphics[width=0.9\linewidth]{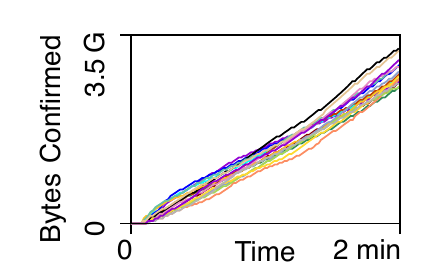}
         \caption{\name}
         \label{fig:geo-progress-new}
     \end{subfigure}%
     \hfill%
     \begin{subfigure}[b]{0.5\linewidth}
         \centering
         \includegraphics[width=0.9\textwidth]{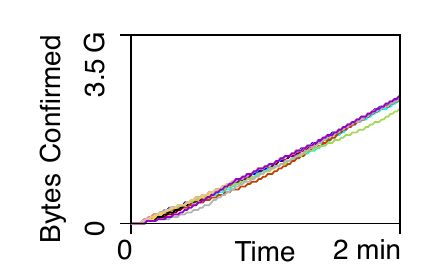}
         \caption{\hb with linking}
         \label{fig:geo-progress-hblinking}
     \end{subfigure}
        \caption{The amount of confirmed data over time when running \name and \hb with inter-node linking on the geo-distributed testbed, plotted on the same scale. Each line represents one server.}
        \label{fig:geo-progress}
\end{figure}

Second, confirmation throughput at different nodes are decoupled, so temporary slowdown at one site will not affect the whole system. Because the system is deployed across the WAN, there are many factors that could cause the confirmation throughput of a node to fluctuate: varying capacity at the network bottleneck, latency jitters, or even behavior of the congestion control algorithm. In \hb, the confirmation progress of all but the $f$ slowest nodes are coupled, so at any time the whole system is only as fast as the $f+1$-slowest node. \name does not have this limitation. Fig.~\ref{fig:geo-progress} shows an example: \name allows each node to always run at its own capacity. 
\hb couples the performance of most servers together, so all servers can only progress at the same, limited rate. \edit{In fact, notice that {\em every} node makes more progress with \name compared to HoneyBadger (with linking) over the 2 minutes shown. The reason is that with \hb, different nodes become the straggler (the $(f+1)^\text{th}$-slowest node) at different time, stalling all other nodes. But with \name, a slow node whose bandwidth improves can accelerate and make progress independently of others, making full use of time periods when it has high bandwidth.} Fig.~\ref{fig:geo-thruput} shows that \name achieves \edit{41\%} better throughput compared to \hb with linking due to this improvement. 

\label{sec:spam-eval}
\edit{Finally, DL-Coupled is 12\% slower than DL on average, but it still achieves 80\% and 23\% higher throughput on average than \hb and \hb with linking. Recall that DL-Coupled constrains nodes that can propose new transactions to prevent spamming attacks. The result shows that in open environments where spamming is a concern, DL-Coupled can still provide significant performance gains. In the rest of the evaluation, we focus on DL (without spam mitigation) to investigate our idea in its purest form.}

\noindent \textbf{Latency.} Confirmation latency is defined as the elapsed time from a transaction entering the system to it being delivered. Similar to the throughput, the confirmation latency at different servers varies due to heterogeneity of the network condition. 
Further, for a particular node, we only calculate the latency of the transactions that this node \textit{itself} generates, i.e. \textit{local} transactions. This is a somewhat artificial metric, but it helps isolate the latency of each server in \hb and makes the results easier to understand. In \hb, a slow node only proposes a new epoch after it has confirmed the previous epoch, so the rate it proposes is coupled with the rate it confirms, i.e. it proposes 1 block after downloading $O(N)$ blocks. Due to this reason, an overloaded node does not have the capacity to even \textit{propose} all the transactions it generates, and whatever transaction it proposes will be stale. When these stale transactions get confirmed at a fast node, the latency (especially the tail latency) at the fast nodes will suffer. Note that \name does not have this problem, because all nodes, even overloaded ones, propose new transactions at a rate limited only by the egress bandwidth. 
In summary, choosing this metric is only advantageous to \hb, so the experiment remains fair. \edit{In Appendix \S\ref{apx:latency-metric}, we provide further details and report the latency of all servers calculated for both local only, and all transactions.}

\begin{figure}
\centering
         \includegraphics[width=\columnwidth]{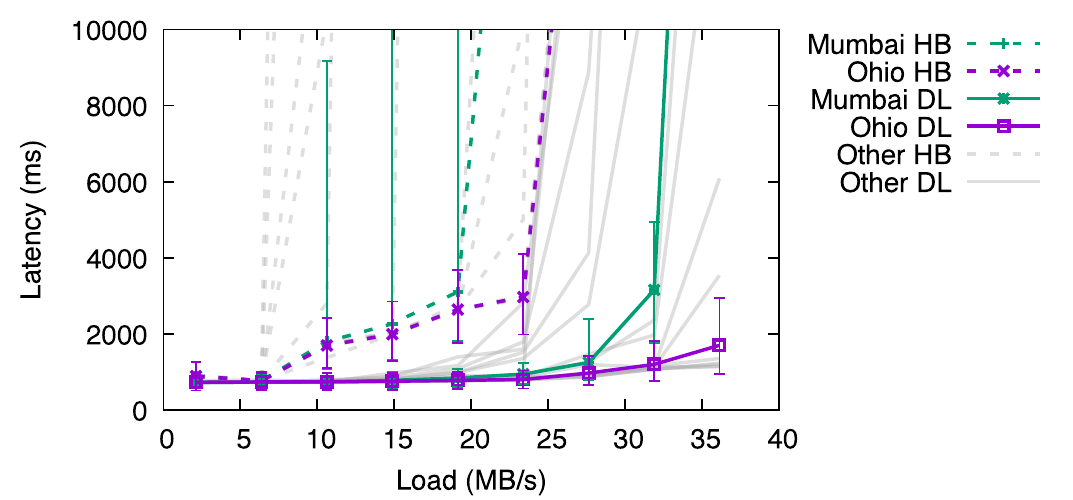}
         \caption{The median latency of \name (solid) and \hb (dash) under different offered load. Error bar shows the 5-th and the 95-th percentiles. Two locations with good (Ohio) and limited (Mumbai) internet connection are highlighted.}
         \label{fig:geo-latency}
\end{figure}

We run the system at different loads and report the latency at each node. In Fig.~\ref{fig:geo-latency}, we focus on two datacenters: Mumbai, which has limited internet connection, and Ohio, which has good internet connection. We first look at the median latency. At low load, both \hb and \name have similarly low median latency. But as we increase the load from 6 MB/s to 15 MB/s, the median latency of \hb increases almost linearly from around 800 ms to 3000 ms. This is because in \hb, proposing and confirming an epoch are done in lockstep. As the load increases, the proposed block becomes larger and takes longer to confirm. This in turn causes more transactions to be queued for the next block so the next proposed block remains large. Actually, the batch (all blocks in an epoch) size of \hb increases from 3.4 MB to 42.5 MB \edit{(200 KB to 2.5 MB per block)} as we increase the load from 6 MB/s to 15 MB/s. Note that \edit{the block size} is not chosen by us, but is naturally found by the system itself. In comparison, the latency of \name only increases by a bit when the load increases, from 730 ms to 830 ms as we increase the load from 2 MB/s to 23 MB/s. \edit{The batch size ranges between 0.85 MB to 11.9 MB (50 KB to 700 KB per block).}

We now look at the tail latency, which is important for service quality. At low load (6 MB/s), the 99-th percentile latency of \name is 1000 ms across all servers, while that of \hb ranges from 1500 ms to 4500 ms. It suggests that \name is more stable. As we increase the load, the tail (95-th percentile) latency of the Mumbai server immediately goes up. This is because \hb does not guarantee all honest blocks to be included in the ledger, and slow nodes are more likely to see their blocks being dropped from an epoch. When it happens, the node has to re-propose the same block in the next epoch, causing significant delay to the block. We note that the tail latency of the Ohio server goes up as well. In comparison, the tail latency of \name at both Mumbai and Ohio stays low until very high load.  

\subsection{Controlled experiments}

\label{sec:controlled-experiments}

In this experiment, we run a series of tests in the controlled setting to verify if \name achieves its design goal: achieving good throughput regardless of network variation. We start 16 servers in one datacenter, and add an artificial one-way propagation delay of 100 ms between each pair of servers to emulate the WAN latency. We then generate synthetic traces for each server that independently caps the ingress and egress bandwidth of the server. For each set of traces, we measure the throughput of \name and \hb.

\begin{figure}
\centering
\begin{subfigure}[b]{0.5\linewidth}
         \centering
         \includegraphics[width=1.0\linewidth]{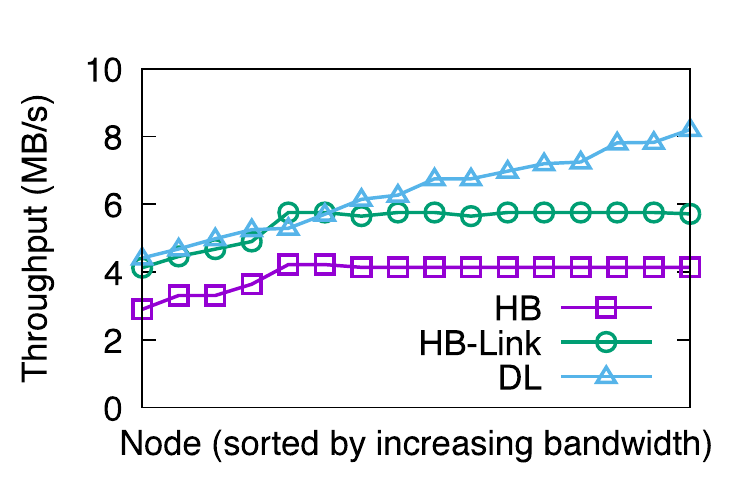}
         \caption{\label{fig:control-thruput-spatial}Spatial variation}%
         
     \end{subfigure}%
     \hfill%
     \begin{subfigure}[b]{0.5\linewidth}
         \centering
         \includegraphics[width=1.0\textwidth]{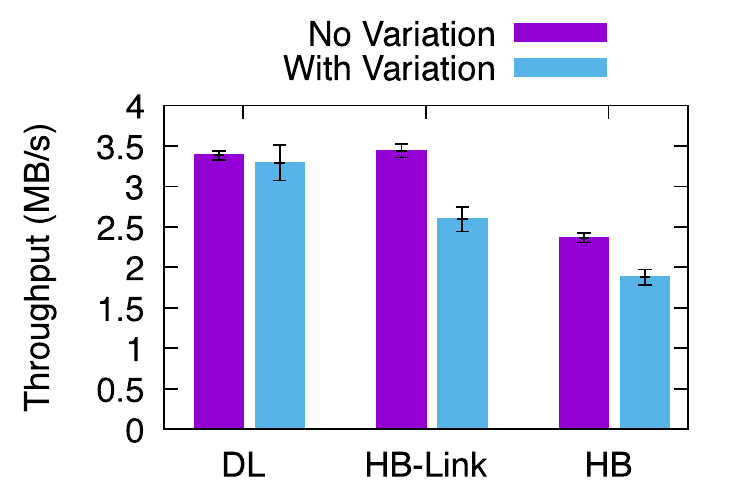}
         \caption{\label{fig:control-thruput-temporal}Temporal variation}%
     \end{subfigure}
        \caption{\label{fig:control-thruput-figures} Throughput of \hb (HB), \hb with linking (HB-Link), and \name (DL) in the controlled experiments. Error bars in (b) show the standard deviation.}%
\end{figure}

\noindent \textbf{Spatial variation.} This is the situation where the bandwidth varies across different nodes but stays the same over time. For the $i$-th node ($0 \leq i < 16$), we set its bandwidth to constantly be $10 + 0.5 i$ MB/s. Fig.~\ref{fig:control-thruput-spatial} shows that the throughput of \hb (with or without linking) is capped at the bandwidth of the fifth slowest server, and the bandwidth available at all faster servers are not utilized. In comparison, the throughput of \name at different servers are fully decoupled. The achieved bandwidth is proportional to the available bandwidth at each server. \name achieves this because it decouples block retrieval at different servers. %

\noindent \textbf{Temporal variation.} We now look at the scenario where the bandwidth varies over time, and show that \name is robust to network fluctuation. We model the bandwidth variation of each node as independent Gauss-Markov processes with mean $b$, variance $\sigma$, and correlation between consecutive samples $\alpha$, and generate synthetic traces for each node by sampling from the process every 1 second. Specifically, we set $b=10$ MB/s, $\sigma=5$ MB/s, $\alpha=0.98$ and generate a trace for each server, i.e. the bandwidth of each server varies independently but have the same distribution with mean bandwidth 10 MB/s. (We show an example of such trace in \S\ref{apx:bwtrace}.) As a comparison, we also run an experiment when the bandwidth at each server does not fluctuate and stays at 10 MB/s. \edit{In our implementation (for all protocols), a node notifies others when it has decoded a block to stop sending more chunks. This optimization is less effective when all nodes have exactly the same fixed bandwidth because all chunks for a block will arrive at roughly the same time. So in this particular experiment, we disable this optimization to enable an apple-to-apple comparison of the fixed and variable bandwidth scenarios.} Fig.~\ref{fig:control-thruput-temporal} shows that as we introduce temporal variation of the network bandwidth, the throughput of \name stays the same. This confirms that \name is robust to network fluctuation. Meanwhile, the throughput of \hb and \hb with linking dropped by 20\% and 25\% respectively.

\subsection{Scalability}
\label{sec:eval-scalability}

In this experiment, we evaluate how \name scales to a large number of servers. As with many evaluations of BFT protocols \cite{hbbft, hotstuff}, we use cluster sizes ranging from 16 to 128.

\begin{figure}[t]
\minipage[t]{0.5\linewidth}
 \centering
         \captionsetup{width=.9\linewidth}
         \includegraphics[width=1.0\textwidth]{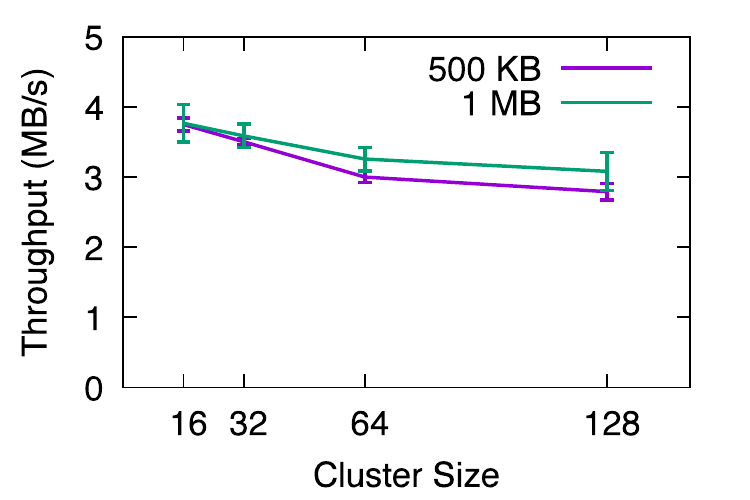}
         \caption{\label{fig:scalability-thruput} Throughput at different cluster size and block size. Error bars show the standard deviation.}
\endminipage\hfill
\minipage[t]{0.5\linewidth}
   \centering
         \captionsetup{width=.9\linewidth}
         \includegraphics[width=1.0\linewidth]{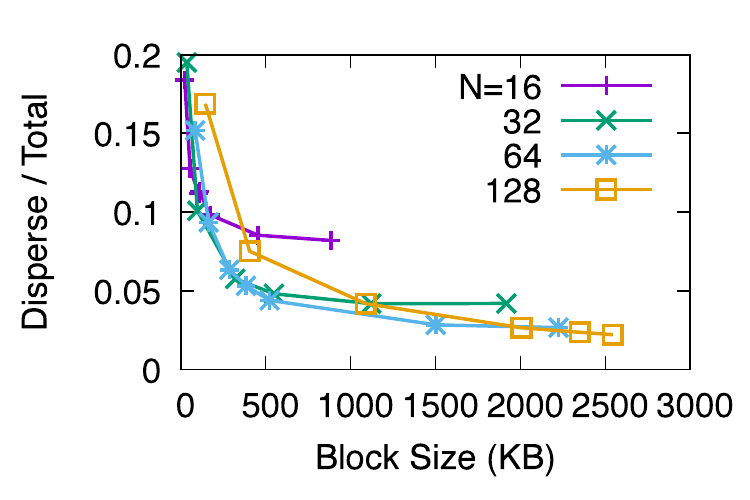}
         \caption{\label{fig:scalability-fraction} Fraction of dispersal traffic versus total traffic at different scale and block size.}
\endminipage
\end{figure}

\noindent \textbf{Throughput.} We first measure the system throughput at different cluster size $N$. For this experiment, we start all the servers in the same datacenter with a 100 ms one-way propagation delay on each link and a 10 MB/s bandwidth cap on each server. We also fix the block size to 500 KB and 1~MB. Fig.~\ref{fig:scalability-thruput} shows that the system throughput slightly drops when $N$ grows 8 times bigger from 16 nodes to 128 nodes. \edit{This is because the BA in the dispersal phase has a per-node cost of $O(N^2)$.} With a constant block size, the messaging overhead takes a larger fraction as $N$ increases. We notice that increasing the block size helps amortize the cost of VID and BA, and results in better system throughput.

\noindent \textbf{Traffic for block dispersal.} A metric core to the design of \name is the amount of data a node has to download in order to participate in block dispersal, i.e. dispersal traffic. More precisely, we are interested in the \edit{\textit{ratio}} of dispersal traffic  to the total traffic (dispersal plus retrieval). The lower this \edit{ratio}, the easier it is for slow nodes to keep up with block dispersal, and the better \name achieves its design goal. Fig.~\ref{fig:scalability-fraction} shows this \edit{ratio} at different scales and block sizes. First, we observe that increasing the block size brings down the fraction of dispersal traffic. This is because a large block size amortizes the fixed cost in VID and BA. Meanwhile, increasing the cluster size reduces the lower bound on the \edit{fraction of dispersal traffic}. This is because in the VID phase, every node is responsible for an $1/(N-2f)$ slice of each block, and increasing $N$ brings down this fraction.

\section{Conclusion}

We presented \name, a new asynchronous BFT protocol that provides near-optimal throughput under fluctuating network bandwidth. \name is based on a novel \edit{restructuring} of BFT protocols that decouples \edit{agreement} from the bandwidth-intensive task of downloading blocks. We implement a full system prototype and evaluate \name on two testbeds across the real internet and a controlled setting with emulated network conditions. Our results on a wide-area deployment across 16 major cities show that \name achieves \edit{2$\times$ better throughput} and 74\% lower latency compared to \edit{\hb}. Our approach \edit{could be} applicable to other BFT protocols, and enables new applications where resilience to poor network condition is vital.

\section*{Acknowledgments}

We would like to thank the National Science Foundation grants CNS-1751009 and CNS-1910676, the Cisco Research Center Award, the Microsoft Faculty Fellowship, and the Fintech@CSAIL program for their support.

\newpage

\bibliographystyle{abbrv}
\bibliography{main}

\appendix
\section{Supplements to the Evaluations}

\subsection{Latency metric}
\label{apx:latency-metric}

\begin{figure}
\begin{center}
     \begin{subfigure}[b]{\linewidth}
         \centering
         \includegraphics[width=0.85\textwidth]{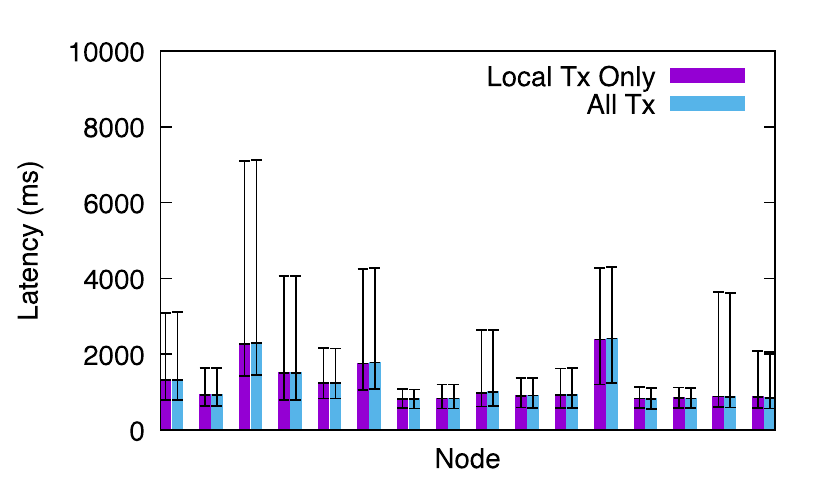}
         \caption{\name}
         \label{fig:latency-metric-new}
     \end{subfigure}
     \begin{subfigure}[b]{\linewidth}
         \centering
         \includegraphics[width=0.85\textwidth]{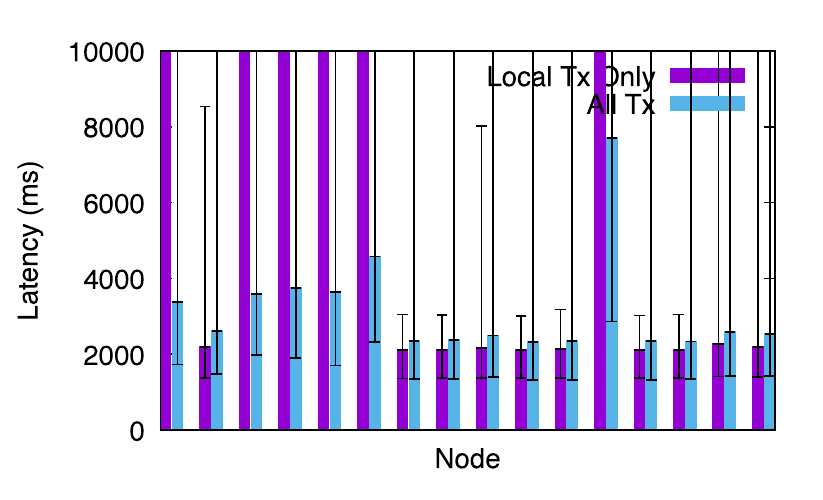}
         \caption{\hb}
         \label{fig:latency-metric-hb}
     \end{subfigure}
        \caption{Confirmation latency of \name and \hb when counting all transactions (All Tx) or only local transactions. Each system runs near its capacity (14.8 MB/s for \hb and 23.4 MB/s for \name). The error bar shows the 5-th and 95-th percentiles.}
        \label{fig:latency-metric}
\end{center}
\end{figure}

Here we justify counting only local transactions when calculating the confirmation latency. As mentioned in \S\ref{sec:internet-performance}, we choose this metric to prevent overloaded servers from impacting the latency (especially the tail latency) of non-overloaded servers. Fig.~\ref{fig:latency-metric} shows the latency of \name and \hb under two metrics: counting all transactions, and counting only local transactions. Each system is running near its capacity. We observe that the latency (both the median and the tail) of \name is the same under the two metrics, so choosing to count only local transactions in no way helps our protocol. For \hb, we observe that by counting all transactions, the median latency of the overloaded servers decreased. This is because the overloaded servers cannot get their local transactions into the ledger (so the local transactions have high latency), but can confirm \textit{some} transactions from other non-overloaded servers. The median latency mostly represents these non-local transactions. Still, these servers are overloaded, and the latency numbers are meaningless because they will increase as system runs for longer. So the latency metric does not matter for the overloaded servers. Meanwhile, we observe that the tail latency of \hb on non-overloaded servers worsens a lot as we switch to counting all transactions. This is due to the transactions proposed by the overloaded nodes, and is the main reason that we choose to count only local transactions. In summary, counting only local transactions for latency calculation does not improve the latency of \name, but helps improve the tail latency of non-overloaded servers in \hb, so choosing this metric is fair.

\subsection{Throughput on another testbed over the internet}
\label{apx:vultr-throughput}

\begin{figure}
\begin{center}
\includegraphics[width=0.85\columnwidth]{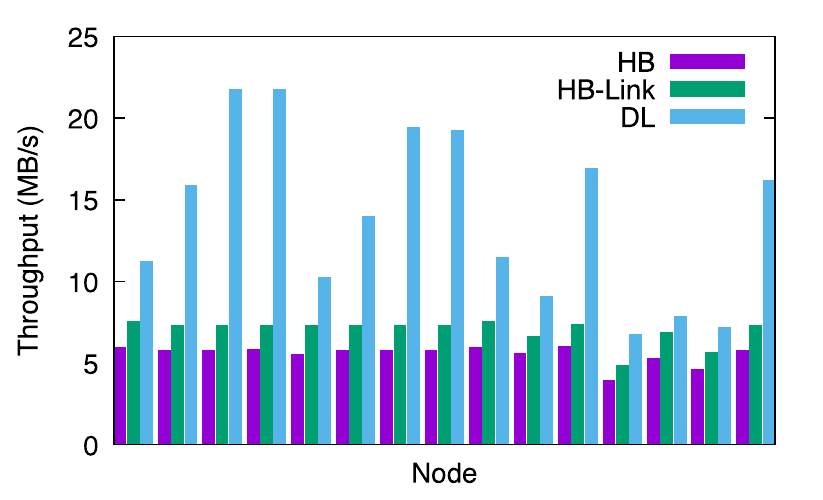}
\end{center}
\caption{\label{fig:geo-thruput-vultr} Throughput of each server running different protocols on the Vultr testbed. HB stands for \hb, HB-Link stands for \hb with inter-node linking, New stands for \name.}
\end{figure}

To further confirm that \name improves the throughput of BFT protocols when running over the internet, we build another testbed on a low-cost cloud provider called Vultr. We use the \$80/mo plan with 6 CPU cores, 16 GB of RAM, 320 GB of SSD, and an 1 Gbps NIC. At the moment of the experiment, Vultr has 15 locations across the globe, and we run one server at each location and perform the same experiment as in \S~\ref{sec:internet-performance}. Fig.~\ref{fig:geo-thruput-vultr} shows the results. \name improves the throughput by at least 50\% over \hb.

\subsection{Example trace of temporal variation}

\label{apx:bwtrace}

\begin{figure}
\begin{center}
\includegraphics[width=0.85\columnwidth]{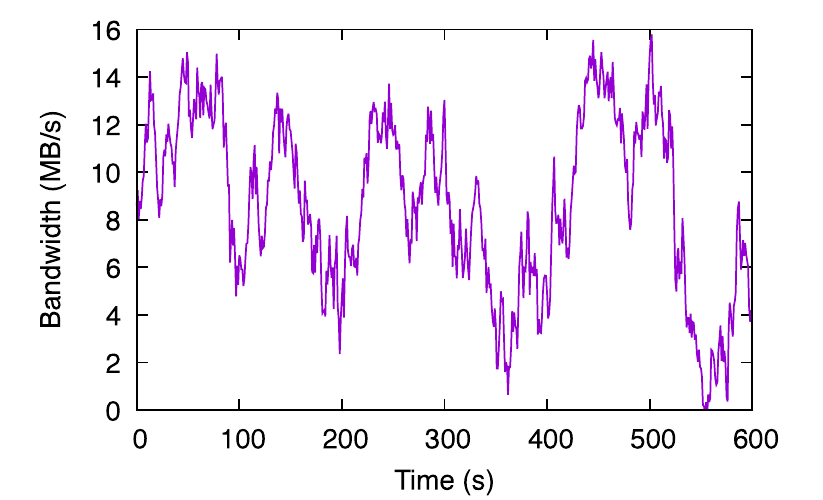}
\end{center}
\caption{\label{fig:bwtrace} A bandwidth trace we used in the temporal variation scenario.}
\end{figure}

We provide in Fig.~\ref{fig:bwtrace} an example of the synthetic bandwidth trace we used in the temporal variation scenario in \S\ref{sec:controlled-experiments}.

\section{\edit{Correctness proof of AVID-M}}
\label{sec:avidm-correctness}

\noindent\textbf{Notations.} We use the symbol ``$\cdot$'' as placeholders in message
parameters to indicate ``any''. For example, $\mathtt{Chunk}(r, \cdot, \cdot)$
means ``$\mathtt{Chunk}$ messages with the first parameter set to $r$ and the other
two parameters set to any value''.

\begin{lemma}
	\label{thm:avid-ready}
	If a correct server sends $\mathtt{Ready}(r)$, then at least one correct
	server has received $N-f$ $\mathtt{GotChunk}(r)$.
\end{lemma}
\begin{proof}
	A correct server broadcasts $\mathtt{Ready}(r)$ in two cases:
	\begin{enumerate}
		\item Having received $N-f$ $\mathtt{GotChunk}(r)$ messages.
		\item Having received $f+1$ $\mathtt{Ready}(r)$ messages.
	\end{enumerate}
	If a correct server sends out $\mathtt{Ready}(r)$ for the aforementioned reason 1,
	then this already satisfies the lemma we want to prove.
	Now assume that a correct server sends $\mathtt{Ready}(r)$ because it has received
	$f+1$ $\mathtt{Ready}(r)$ (the aforementioned reason 2).
	Then there must exist a correct server which has sent out $\mathtt{Ready}(r)$
	because of the aforementioned reason 1. Otherwise, there can be at most
	$f$ $\mathtt{Ready}(r)$ messages (forged by the $f$ Byzantine servers),
	and no correct server will ever send $\mathtt{Ready}(r)$ because of reason 2,
	which contradicts with our assumption. So there exists a correct server that
	has received $N-f$ $\mathtt{GotChunk}(r)$, and this satisfies the lemma.
\end{proof}

\begin{theorem}[Termination]
	\label{thm:avid-termination}
	If a correct client invokes $\mathtt{Disperse}$ and no other client invokes
	$\mathtt{Disperse}$ on the same instance of VID, then all correct
	servers eventually $\mathtt{Complete}$ the dispersal.
\end{theorem}
\begin{proof}
	A correct client sends correctly encoded chunks to all servers.
	Let's assume the Merkle root of the chunks is $r$, then all correct
	servers eventually receive $\mathtt{Chunk}(r, \cdot, \cdot)$.
	Because there is no other client invoking $\mathtt{Disperse}$,
	it is impossible for a server to receive $\mathtt{Chunk}(r', \cdot, \cdot)$ for any $r' \ne r$,
	and no correct server will ever broadcast $\mathtt{GotChunk}(r')$ for any $r' \ne r$.
	So each correct server will send out $\mathtt{GotChunk}(r)$. Eventually, all correct servers
	will receive $N-f$ $\mathtt{GotChunk}(r)$.

    All correct servers will broadcast
	$\mathtt{Ready}(r)$ upon receiving these $N-f$ $\mathtt{GotChunk}(r)$
	messages or they have already sent $\mathtt{Ready}(r)$. A correct server will
	$\mathtt{Complete}$ upon receiving $2f+1$ $\mathtt{Ready}(r)$. We have shown
	that all $N-f$ correct servers will eventually send $\mathtt{Ready}(r)$.
	Because $N-f \ge 2f+1$, all correct servers will $\mathtt{Complete}$.
\end{proof}

\begin{lemma}
	\label{thm:avid-unique-ready}
	If a correct server has sent out $\mathtt{Ready}(r)$, then \textit{no} correct server will ever send out $\mathtt{Ready}(r')$ for any $r' \ne r$.
\end{lemma}
\begin{proof}
	Let's assume for contradiction that two messages
	$\mathtt{Ready}(r)$ and $\mathtt{Ready}(r')$ ($r \ne r'$) have both been sent by correct servers.
	By Lemma \ref{thm:avid-ready}, at least one correct server has received
	$N-f$ $\mathtt{GotChunk}(r)$, and at least one correct server has received
	$N-f$ $\mathtt{GotChunk}(r')$ ($r' \ne r$).

	We obtain a contradiction by showing that the system cannot generate
	$N-f$ $\mathtt{GotChunk}(r)$ messages plus $N-f$ $\mathtt{GotChunk}(r')$ messages
	for the two correct servers to receive.
	Assume $h$ $\mathtt{GotChunk}(r)$ messages come from correct servers, $h'$
	$\mathtt{GotChunk}(r')$ come from correct servers, and there are $\beta$
	Byzantine servers ($\beta \le f$ by the definition of $f$).
	Then we have
	\begin{align*}
		h + \beta & \ge N - f \\
		h' + \beta & \ge N - f.
	\end{align*}
	A correct server do not broadcast \textit{both} $\mathtt{GotChunk}(r)$ and $\mathtt{GotChunk}(r')$,
	while a Byzantine server is free to send
	different $\mathtt{GotChunk}$ messages to different correct servers,
	so we have
	\begin{align*}
		h + h' + \beta & \le N.
	\end{align*}
	These constraints imply
	\begin{align*}
		\beta \ge N-2f.
	\end{align*}
	However, $\beta \le f$, so we must have $N \le 3f$. This contradicts with
	our assumption of $N \ge 3f + 1$ in our security model (\S\ref{sec:model}),
	so it is impossible, and the assumption must not hold.
\end{proof}

\begin{theorem}[Agreement]
	\label{thm:avid-agreement}
	If some correct server $\mathtt{Complete}$s the dispersal, then all correct servers will
	eventually $\mathtt{Complete}$ the dispersal.
\end{theorem}
\begin{proof}
	A correct server $\mathtt{Complete}$s if and only if it has received $2f+1$
	$\mathtt{Ready}(r)$ messages. We want to prove that in this situation, all correct servers
	will eventually \textit{send} a $\mathtt{Ready}(r)$, so that they will all
	\textit{receive} at least $2f+1$ $\mathtt{Ready}(r)$ messages needed to
	$\mathtt{Complete}$.

	We now assume a correct server has $\mathtt{Complete}$d after receiving
	$2f+1$ $\mathtt{Ready}(r)$. Out of these messages,
	at least $f+1$ must be broadcast from correct servers, so all correct servers
	will eventually receive these $\mathtt{Ready}(r)$. A correct server will send out
	$\mathtt{Ready}(r)$ upon receiving $f+1$ $\mathtt{Ready}(r)$, so all correct servers
	will do so upon receiving the aforementioend $f+1$ $\mathtt{Ready}(r)$ messages.

	Because all correct servers will send $\mathtt{Ready}(r)$, eventually
	all correct servers will receive $N-f$ $\mathtt{Ready}(r)$.
	Because $N-f \ge 2f+1$, all of them will $\mathtt{Complete}$.
\end{proof}

\begin{lemma}
	\label{thm:avid-agree-root}
	If a correct server has $\mathtt{Complete}$d, then all correct servers eventually
	set the variable $\mathtt{ChunkRoot}$ to the same value.
\end{lemma}
\begin{proof}
	A correct server uses $\mathtt{ChunkRoot}$ to store the root of the chunks of
	the dispersed block, so we are essentially proving that all correct servers
	agree on this root. Assume that a server $\mathtt{Complete}$s, then it must have
	received $2f+1$ $\mathtt{Ready}(r)$ messages. We now prove that no correct
	server can ever receive $2f+1$ $\mathtt{Ready}(r')$ messages for any $r' \ne r$.
	Because a correct server has received $2f+1$ $\mathtt{Ready}(r)$, there must
	be $f+1$ correct servers who have broadcast $\mathtt{Ready}(r)$.
	By Lemma \ref{thm:avid-unique-ready}, no correct server will ever broadcast
	$\mathtt{Ready}(r')$ for any $r' \ne r$, so a correct server can receive
	at most $f$ $\mathtt{Ready}(r')$ for any $r' \ne r$, which are forged by the $f$
	Byzantine servers.

	By Theorem \ref{thm:avid-agreement}, all correct servers eventually $\mathtt{Complete}$,
	so they must eventually receive $2f+1$ $\mathtt{Ready}(r)$, and will each set
	$\mathtt{ChunkRoot} = r$.
\end{proof}

\begin{theorem}[Availability]
	\label{thm:avid-availability}
	If a correct server has $\mathtt{Complete}$d, and a correct client invokes
	$\mathtt{Retrieve}$, it eventually reconstructs some block $B'$.
\end{theorem}
\begin{proof}
	The $\mathtt{Retrieve}$ routine returns at a correct client as long as it can collect
	$N-2f$ $\mathtt{ReturnChunk}(r, C_i, P_i)$ messages with the same root $r$
	and valid proofs $P_i$. A correct server sends $\mathtt{ReturnChunk}(\mathtt{MyRoot}, \mathtt{MyChunk}, \mathtt{MyProof})$
	to a client as long as it has $\mathtt{MyRoot}$, $\mathtt{MyChunk}$, $\mathtt{MyProof}$, and
	$\mathtt{ChunkRoot}$ set, and $\mathtt{MyRoot} = \mathtt{ChunkRoot}$.
	Here, a server uses $\mathtt{MyRoot}$ to store the root of the chunk it has received,
	uses $\mathtt{MyChunk}$ to store the chunk, and uses $\mathtt{MyProof}$ to store the Merkle proof (Fig.~\ref{fig:dispersalAlgo}).
	We now prove that if any correct server $\mathtt{Complete}$s, at least $N-2f$ correct
	servers will eventually meet this condition and send $\mathtt{ReturnChunk}$ to the client.

	Assume that a correct server has $\mathtt{Complete}$d
	the VID instance with $\mathtt{ChunkRoot}$ set to $r$.
	Then, by Lemmas \ref{thm:avid-agreement}, \ref{thm:avid-agree-root},
	all correct servers will eventually $\mathtt{Complete}$ and set
	$\mathtt{ChunkRoot} = r$.
	Also, this server must have received $2f+1$ $\mathtt{Ready}(r)$
	messages, out of which at least $f+1$ must come from correct servers.
	According to Lemma \ref{thm:avid-ready}, at least one correct server has
	received $N-f$ $\mathtt{GotChunk}(r)$. At least $N-2f$ $\mathtt{GotChunk}(r)$
	messages must come from correct servers, so they each must have
	$\mathtt{MyChunk}$, $\mathtt{MyProof}$ set, and have set $\mathtt{MyRoot} = r$.

	We have proved that at least $N-2f$ correct servers will send
	$\mathtt{ReturnChunk}(r, C_i, P_i)$ messages. For each message sent
	by the $i$-th server (which is correct), $P_i$ must be a valid proof showing
	$C_i$ is the $i$-th chunk under root $r$, because the server has validated this proof.
	So the client will eventually obtain the $N-2f$ chunks needed to reconstruct a block.
\end{proof}

\begin{lemma}
	\label{thm:avid-client-agree}
	Any two correct clients finishing $\mathtt{Retrieve}$ have their variable $\mathtt{ChunkRoot}$
	set to the same value.
\end{lemma}
\begin{proof}
	A client uses variable $\mathtt{ChunkRoot}$ to store the root of the $N-2f$ chunks it uses
	to reconstruct the block (Fig.~\ref{fig:retrievalAlgo}), so we are essentially proving
	that any two correct clients will use chunks under the same root when
	executing $\mathtt{Retrieve}$. Let's assume for contradiction that two correct
	clients finish $\mathtt{Retrieve}$, but have set $\mathtt{ChunkRoot}$ to $r$ and $r'$
	respectively ($r \ne r'$). This implies that one client has received at least $N-2f$
	$\mathtt{ReturnChunk}(r, \cdot, \cdot)$ messages, and the other has received $N-2f$
	$\mathtt{ReturnChunk}(r', \cdot, \cdot)$ messages. Out of these messages, at least
	$N-3f$ $\mathtt{ReturnChunk}(r, \cdot, \cdot)$ and at least $N-3f$
	$\mathtt{ReturnChunk}(r', \cdot, \cdot)$ are from correct servers (because $N \ge
	3f+1$ by our security assumptions in \S\ref{sec:model}). Since a correct server
	ensures $\mathtt{MyRoot} = \mathtt{ChunkRoot}$ and uses
	$\mathtt{MyRoot}$ as the first parameter of $\mathtt{ReturnChunk}$ messages,
	there must exist some correct server with $\mathtt{ChunkRoot}$ set to $r$, and
	some correct server with $\mathtt{ChunkRoot}$ set to $r'$.
	Also, since a correct server only sends $\mathtt{ReturnChunk}$ when it has
	$\mathtt{Complete}$d, there must be some server which has $\mathtt{Complete}$d.
	This contradicts with Lemma \ref{thm:avid-agree-root}, which states that all
	correct servers must have $\mathtt{ChunkRoot}$ set to the same value. The assumption
	must not hold.
\end{proof}

\noindent\textbf{Extra notations.}
To introduce the following lemma, we need to define a few extra notations.
Let $\mathtt{Encode}(B)$ be the encoding result of a block $B$ in the form
of an array of $N$ chunks. Let $\mathtt{Decode}(C)$ be the decoding result (a block)
of an array of $N-2f$ chunks. Let $\mathtt{MerkleRoot}(C)$ be the Merkle root
of an array of chunks.

\begin{lemma}
	\label{thm:avid-root-summary}
	For any array of $N$ chunks $C$, exactly one of the following is true:
	\begin{enumerate}
		\item For any two subsets $D_1$, $D_2$ of $N-2f$ chunks in $C$,
			$\mathtt{Decode}(D_1) = \mathtt{Decode}(D_2)$.
		\item For any subset $D$ of $N-2f$ chunks in $C$, $\mathtt{MerkleRoot}( \mathtt{Encode}( \mathtt{Decode}( D ) ) ) \ne \mathtt{MerkleRoot}(C)$.
	\end{enumerate}
\end{lemma}
\begin{proof}
	We are proving that a set of chunks $C$ is either:
	\begin{enumerate}
		\item Correctly encoded (consistent), so any subset of $N-2f$ chunks in $C$ decode into the same block.
		\item Or, no matter which subset of $N-2f$ chunks in $C$ are used for decoding, a correct client can re-encode the decoded block, compute the Merkle root over the encoding result,
	and find it to be \textit{different} from the Merkle root of $C$, and thus detect an encoding error.
	\end{enumerate} 

	\textit{Case 1: Consistent encoding.} Assume for any subset $D$ of $N-2f$
	chunks in $C$, $\mathtt{Decode}(D) = B$. We now want to prove that
	$\mathtt{MerkleRoot}( \mathtt{Encode}( \mathtt{Decode}( D ) ) ) = \mathtt{MerkleRoot}(C)$.
	By our assumption, $\mathtt{Encode}( \mathtt{Decode}( D ) ) = \mathtt{Encode}(B)$,
	so we only need to show $C = \mathtt{Encode}(B)$. This is clearly true by
	the definition of erasure code: the $\mathtt{Encode}$ function encodes $B$ into
	a set of $N$ chunks, of which any subset of $N-2f$ chunks will decode into $B$.
	$C$ already satisfies this property, and
	the $\mathtt{Encode}$ process is deterministic, so it must be $\mathtt{Encode}(B) = C$,
	and the lemma is satisfied in this case.

	\textit{Case 2: Inconsistent encoding.} Assume there exist two subsets $D_1$, $D_2$
	of $N-2f$ chunks in $C$, and $\mathtt{Decode}(D_1) \ne \mathtt{Decode}(D_2)$.
	Let $\mathtt{Decode}(D_1) = B_1$ and $\mathtt{Decode}(D_2) = B_2$ where $B_1 \ne B_2$.
	We want to prove that for any subset $D$ of $N-2f$ chunks in $C$,
	$\mathtt{MerkleRoot}( \mathtt{Encode}( \mathtt{Decode}( D ) ) ) \ne \mathtt{MerkleRoot}(C)$.
	
	We prove it by showing there does \textit{not} exist any block $B$ such that
	$C = \mathtt{Encode}(B)$. That is, $C$ is not a consistent encoding result of any
	block. Assume for contradiction that there exists $B'$ such that
	$C = \mathtt{Encode}(B')$. Because $D_1$ is a subset of $N-2f$ chunks in $C$ and
	$\mathtt{Decode}(D_1) = B_1$, it must be $B_1 = B'$, otherwise the semantic of
	erasure code is broken. For the same reason $B_2 = B'$, so $B_1 = B_2$. However
	it contradicts with $B_1 \ne B_2$, so the assumption must not hold, and there does
	not exist any block $B$ such that $C = \mathtt{Encode}(B)$.

	We now prove that
	$\mathtt{MerkleRoot}( \mathtt{Encode}( \mathtt{Decode}( D ) ) ) \ne \mathtt{MerkleRoot}(C)$
	for any subset $D$ of $N-2f$ chunks in $C$. Assume for contradiction that
	$\mathtt{MerkleRoot}( \mathtt{Encode}( \mathtt{Decode}( D ) ) ) = \mathtt{MerkleRoot}(C)$,
	then it must be that $C = \mathtt{Encode}( \mathtt{Decode}( D ) )$
	because Merkle root is a secure summary of the chunks. This contradicts with the result
	we have just proved: there does not exist any block $B$ such that
	$C = \mathtt{Encode}(B)$. So the assumption cannot hold, and the lemma is satisfied in this case.
\end{proof}

\begin{theorem}[Correctness]
	\label{thm:avid-correctness}
	If a correct server has $\mathtt{Complete}$d, then correct clients always reconstruct the \textit{same} block $B'$ by invoking $\mathtt{Retrieve}$. Also, if a correct client initiated the dispersal by invoking $\mathtt{Disperse}(B)$ and no other client invokes $\mathtt{Disperse}$, then $B=B'$.
\end{theorem}
\begin{proof}
	We first prove the first half of the theorem: any two correct clients always return the same
	data upon finishing $\mathtt{Retrieve}$. By Lemma \ref{thm:avid-client-agree}, any two
	clients will set their $\mathtt{ChunkRoot}$ to the same value. Note that a client sets
	$\mathtt{ChunkRoot}$ to the root of the chunks it uses for decoding. This implies that
	any two correct clients will use subsets from the \textit{same} set of
	chunks. By Lemma \ref{thm:avid-root-summary}, either:
	\begin{enumerate}
		\item They both decode and obtain the same block $B'$.
		\item Or, they each compute $\mathtt{MerkleRoot}( \mathtt{Encode}())$ on the decoded
	block and both get a result that is different from $\mathtt{ChunkRoot}$.
	\end{enumerate}
	In the first situation, both clients will return $B'$. In the second situation, they both
	return the block containing string ``BAD\_UPLOADER''. In either case, they return the same block.

	We then prove the second half of the theorem. Assume a correct client has initiated
	$\mathtt{Disperse}(B)$ and no other client invokes $\mathtt{Disperse}$. By
	Theorem \ref{thm:avid-availability}, any correct client invoking $\mathtt{Retrieve}$
	will obtain some block $B'$. We now prove that $B' = B$.
	Assume for contradiction that $B' \ne B$. Then the client must have received
	$N-2f$ $\mathtt{ReturnChunk}(\mathtt{MerkleRoot}(\mathtt{Encode}(B')), \cdot, \cdot)$
	messages. At least one of them must come from a correct server because $N-2f > f$, so
	at least one correct server have $\mathtt{ChunkRoot}$ set to
	$\mathtt{MerkleRoot}(\mathtt{Encode}(B'))$. However, because there is only
	invocation of $\mathtt{Disperse}(B)$, all correct servers must have
	set $\mathtt{ChunkRoot}$ to $\mathtt{MerkleRoot}(\mathtt{Encode}(B))$.
	So $\mathtt{MerkleRoot}(\mathtt{Encode}(B)) = \mathtt{MerkleRoot}(\mathtt{Encode}(B'))$
	This contradicts with our assumption, so the assumption must not hold, and $B = B'$.
\end{proof}

\section{Specification of the full DispersedLedger protocol with Inter-node Linking}
\label{sec:appendix-internodeLinking}

Figure~\ref{fig:internodeLinkingAlgo} describes how to modify the single-epoch protocol to use inter-node linking. Blue color highlights the parts are added compared to the single-epoch protocol.

\begin{figure}
\centering
\fbox{%
  \parbox{0.95\columnwidth}{%

\underline{\textbf{Phase 1. Dispersal} at the $i$-th server}
\begin{enumerate}
	\item \blue{For $1 \le j \le N$, let $V^e_i[j]$ be the largest epoch number $t$ such that
		$\text{VID}^1_j, \text{VID}^2_j, \dots, \text{VID}^t_j$ have $\mathtt{Complete}$d.}
	\item Let $B_i^e$ be the block to disperse (propose) for epoch $e$. \blue{$B_i^e$ contains two
		parts: transactions $T_i^e$ and observation $V^e_i$.}
    \item Invoke $\mathtt{Disperse}(B_i^e)$ on $\text{VID}_i^e$ as a client.
\end{enumerate}
\begin{itemize}
	\item Upon $\mathtt{Complete}$ of $\text{VID}_j^e$ ($1 \le j \le N$), if we have not invoked $\mathtt{Input}$ on $\text{BA}_j^e$, invoke $\mathtt{Input}(1)$ on  $\text{BA}_j^e$.
	\item Upon $\mathtt{Output}(1)$ of least $N-f$ BA instances, invoke $\mathtt{Input}(0)$ on all remaining BA instances on which we have not invoked $\mathtt{Input}$.
	\item Upon $\mathtt{Output}$ of all BA instances,
		\begin{enumerate}
			\item Let local variable $S^e_i \subset \{1 \dots N\}$ be the indices of all BA instances that $\mathtt{Output}(1)$. That is, $j \in S$ if and only if $\text{BA}_j^e$ has $\mathtt{Output}(1)$ at the $i$-th server.
			\item Move to retrieval phase.
		\end{enumerate}
\end{itemize}

\underline{\textbf{Phase 2. Retrieval}}

\begin{enumerate}
	\item For all $j \in S^e_i$, invoke $\mathtt{Retrieve}$ on $\text{VID}_j^e$ to download full block ${B_j^e}'$. \blue{Decompose ${B_j^e}'$ into transactions ${T_j^e}'$ and observation ${V^e_j}'$. Let
		${V^e_j}' = [\infty, \infty, \dots, \infty]$ if ${B_j^e}'$ is ill-formatted.}
	\item Deliver $\{{T_j^e}' | j \in S^e_i\}$ (sorted by increasing indices). \blue{Set $\mathtt{Delivered}[e][j] = 1$ (initially $0$) for all $j \in S^e_i$.}
    \item \blue{For $1 \le j \le n$, let $E^e_i[j]$ be the $(f+1)^{th}$-largest value among
	    $\{{V_k^e}'[j] | k \in S^e_i\}$.}
    \item \blue{For all $1 \le j \le N$, for all $1 \le d \le E^e_i[j]$, check if
	    $\mathtt{Delivered}[d][j] = 0$. If so, invoke $\mathtt{Retrieve}$
		on $\text{VID}^d_j$ to download full block ${B_j^d}'$, and
		set $\mathtt{Delivered}[d][j] = 1$ (initially $0$).}
	\item \blue{Deliver all blocks downloaded in step 4 (sorted by increasing epoch number
		and node index).}
\end{enumerate}

} }
\caption{
Algorithm for DispersedLedger with inter-node linking. The blue color indicates the changes from the single-epoch algorithm.
}
\label{fig:internodeLinkingAlgo}

\end{figure}

\section{Correctness proof of \name}
\label{sec:dlproof}

\noindent\textbf{Notations.} Let $H$ ($H \subset \{1, 2, \dots, N\}$) be the set of the indices
of correct nodes. That is, $i \in H$ if and only if the $i$-th node is correct. In our proof, we
use the variables in the full algorithm defined in Fig.~\ref{fig:internodeLinkingAlgo}. We also
use ``phase $x$, step $y$'' to refer to specific steps
in Fig.~\ref{fig:internodeLinkingAlgo}.

\begin{lemma}
	\label{thm:ba-availability} For any epoch $e$, any $i \in H$, and any $1 \le j \le N$,
	if $j \in S^e_i$ then $\text{VID}^e_i$ has $\mathtt{Complete}$d at some correct node.
\end{lemma}
\begin{proof}
	By the definition of $S^e_i$ (phase 1, step 3), $j \in S^e_i$ if and only if
	$\text{BA}^e_j$ has $\mathtt{Output}(1)$ at the $i$-th node.
	By the Validity property of BA (\S\ref{sec:ba-def}),
	$\text{BA}^e_j$ $\mathtt{Output}(1)$ at a correct node implies that at least
	one correct node has invoked $\mathtt{Input}(1)$ on $\text{BA}^e_j$,
	which only happens when that node sees $\text{VID}^e_j$ $\mathtt{Complete}$
	(phase 1, step 3).
\end{proof}

\begin{lemma}
	\label{thm:counter-agreement} For any epoch $e$, any $i, j \in H$, $S^e_i = S^e_j$ and
	$E^e_i = E^e_j$.
\end{lemma}
\begin{proof}
	By the definition of $S^e_i$ (phase 1, step 3), $k \in S^e_i$ if and only if $\text{BA}^e_k$ has
	$\mathtt{Output}(1)$ at the $i$-th node. By the Agreement property of BA (\S\ref{sec:ba-def}),
	$\text{BA}^e_k$ will eventually $\mathtt{Output}(1)$ at the $j$-th node. So
	$k \in S^e_i$ if and only if $k \in S^e_j$, and $S^e_i = S^e_j$.

	We now prove $E^e_i = E^e_j$. The $i$-th node (which is correct) starts the
	computation of $E^e_i$
	by invoking $\mathtt{Retrieve}$ on all VIDs in $\{\text{VID}^e_k | k \in S^e_i\}$.
	These $\mathtt{Retrieve}$s are guaranteed to finish by Lemma \ref{thm:ba-availability}
	and the Availability property of VID (Theorem \ref{thm:avid-availability}).
	The node then extracts the observations $\{{V^e_k}' | k \in S^e_i\}$ from the downloaded
	blocks. Note that the $j$-th node will download the same set of observations.
	This is because $S^e_i = S^e_j$, and the VID Correctness property (Theorem \ref{thm:avid-correctness})
	guarantees the $j$-th node will obtain the same blocks when invoking $\mathtt{Retrieve}$
	on $\{\text{VID}^e_k | k \in S^e_j\}$.

	To combine the observations into the estimation, the $i$-th node runs phase 2, step 3.
	This process is deterministic, with $E^e_i$ being a function of the observations
	$\{{V^e_k}' | k \in S^e_i\}$ and parameter $f$. Because we have just proved the $j$-th
	node will obtain the same set of estimations, and by our security model $f$
	is a protocol parameter known to all nodes (\S\ref{sec:model}), the $j$-th node
	will get the same results.

\end{proof}

\begin{lemma}
	\label{thm:ba-progress} For any epoch $e$, and any $i \in H$, $|S^e_i| \ge N-f$. That is,
	$S^e_i$ contains at least $N-f$ indices.
\end{lemma}
\begin{proof}
	By the definition of $S^e_i$ (phase 1, step 3), this lemma essentially states that 
	at least $N-f$ BAs among $\{\text{BA}^e_1, \text{BA}^e_2, \dots, \text{BA}^e_N\}$
	will $\mathtt{Output}(1)$ at the $i$-th node.

	Assume for contradiction that $|S^e_i| < N-f$. By Lemma \ref{thm:counter-agreement},
	$|S^e_j| < N-f$ for all $j \in H$, i.e., less than $N-f$ BAs eventually $\mathtt{Output}(1)$
	at every correct node. We now consider the possible outcomes of the remaining BA instances,
	which do not eventually $\mathtt{Output}(1)$.

	One possibility is some of them $\mathtt{Output}(0)$. According to phase 1, step 3, correct
	nodes will not invoke $\mathtt{Input}(0)$ on any BA instance unless $N-f$ BA instances
	have $\mathtt{Output}(1)$. By our assumption, less than $N-f$ BA $\mathtt{Output}(1)$,
	so the latter is not happening and \textit{no} correct nodes will $\mathtt{Input}(0)$
	on \textit{any} BA instance. By the Validity property of BA (\S\ref{sec:ba-def}),
	no BA instance can $\mathtt{Output}(0)$.

	We have showed that the remaining BAs cannot $\mathtt{Output}(0)$, so it must be
	that all of them never terminate. We will prove it is also impossible. Assume
	for contradication that all BAs that do not $\mathtt{Output}(1)$ never terminate.
	By our assumption, less than $N-f$ BAs $\mathtt{Output}(1)$, so there must exist
	$k \in H$ such that $\text{BA}^e_k$ never terminates. By the Termination
	property of VID (Theorem \ref{thm:avid-termination}), $\text{VID}^e_k$ eventually
	$\mathtt{Complete}$s on all correct nodes. According to phase 1, step 3,
	because not all BAs will terminate, all correct nodes will stay at this step.
	All correct nodes will $\mathtt{Input}(1)$ to $\text{BA}^e_k$ upon seeing
	$\text{VID}^e_k$ $\mathtt{Complete}$. By the Termination and Validity properties
	of BA (\S\ref{sec:ba-def}), $\text{BA}^e_k$ will terminate and $\mathtt{Output}(1)$,
	which conflicts with our assumption.

	We have showed there is no valid outcome for the remaining BA instances, so our
	assumption cannot hold, and at least $N-f$ BA instances eventually $\mathtt{Output}(1)$
	at all correct nodes.
\end{proof}

\begin{lemma}
	\label{thm:correct-estimation} For any epoch $e$, any $i \in H$, and any $1 \le j \le N$, there exist
	$p, q \in H$ such that $V^e_p[j] \le E^e_i[j] \le V^e_q[j]$.
\end{lemma}
\begin{proof}
	The lemma states that if the $i$-th node (which is correct) computes the estimation $E^e_i[j]$
	for the $j$-th node, then the estimation is lower- and upper-bounded by the
	\textit{observations} $V^e_p[j]$ and $V^e_q[j]$ of two correct nodes (with indices
	$p$ and $q$). That is, the estimation is not too high or too low.

	Now assume for contradication that for some $1 \le j \le N$, for all $p \in H$,
	$V^e_p[j] > E^e_i[j]$. That is, the estimation for $j$ is not lower bounded by
	the observations made by any correct node.
	According to phase 2, step 3, the $i$-th node sets $E^e_i[j]$ to the
	$(f+1)^{th}$-largest value among $\{{V^e_k}'[j] | k \in S^e_i\}$.
	Here, ${V^e_k}'$ is the observation of the $k$-th node downloaded by invoking
	$\mathtt{Retrieve}$ on $\text{VID}^e_k$. By Lemma \ref{thm:ba-availability} and
	VID Availability property (Theorem \ref{thm:avid-availability}), the $\mathtt{Retrieve}$s
	will eventually finish.

	By our assumption, for all $p \in H \cap S^e_i$, $V^e_p[j] > E^e_i[j]$.
	By the VID Correctness property (Theorem
	\ref{thm:avid-correctness}), the observations
	of correct nodes will be correctly downloaded.
	That is, ${V^e_k}' = V^e_k$ for all $k \in H$. So
	for all $p \in H \cap S^e_i$, ${V^e_p}'[j] > E^e_i[j]$.
	By Lemma \ref{thm:ba-progress}, $|S^e_i| \ge N-f$, so $|H \cap S^e_i| \ge N-2f$.
	So there are \textit{at least} $N-2f$ values in $\{{V^e_k}'[j] | k \in S^e_i\}$
	that are greater than $E^e_i[j]$. However, $E^e_i[j]$ is the
	$(f+1)^{th}$-largest value among $\{{V^e_k}'[j] | k \in S^e_i\}$,
	so there can be \textit{at most} $f$ values in $\{{V^e_k}'[j] | k \in S^e_i\}$
	that are greater than $E^e_i[j]$. Because $N>3f$ (\S\ref{sec:model}), $N-2f > f$,
	so the two conclusions are in conflict, and the assumption cannot hold.

	We can similarly prove it is impossible that for some $1 \le j \le N$, for all $q \in H$,
	$V^e_q[j] < E^e_i[j]$.
\end{proof}

\begin{theorem}[\name is well-defined]
	\label{thm:progress} For any epoch $e$, any $i \in H$, the $i$-th node eventually finishes epoch $e$.
\end{theorem}
\begin{proof}
	This lemma states that correct nodes will never be stuck in any epoch $e$, so that
	our algorithm is well-defined. To prove that,
	we go through Fig. \ref{fig:internodeLinkingAlgo} line by line and prove each step will
	eventually finish.

	\textit{Phase 1, steps 1--2.} These are local computation and will finish instantly.

	\textit{Phase 1, step 3.} This step finishes as soon as all BA instances in that epoch
	$\mathtt{Output}$. By Lemma \ref{thm:ba-progress}, all correct nodes eventually see
	at least $N-f$ BA instances $\mathtt{Output}(1)$. At that point, each correct node will
	invoke $\mathtt{Input}(0)$ into all BAs on which it has not invoked $\mathtt{Input}$.
	This ensures that all correct nodes eventually invoke $\mathtt{Input}$ on all BAs.
	By the Termination property of BA (\S\ref{sec:ba-def}), all BAs will eventually
	$\mathtt{Output}$ on all correct nodes, which ensures this step will finish.

	\textit{Phase 2, step 1.} This step finishes as soon as $\mathtt{Retrieve}$s on
	$\{\text{VID}^e_j | j \in S^e_i\}$ finish. By Lemma \ref{thm:ba-availability},
	$\{\text{VID}^e_j | j \in S^e_i\}$ will $\mathtt{Complete}$ on all correct nodes.
	Then by VID Availability property (Theorem \ref{thm:avid-availability}),
	the $\mathtt{Retrieve}$s will finish.

	\textit{Phase 2, steps 2--3.} These are local computation and will finish instantly.

	\textit{Phase 2, step 4.} This step will finish if for all $1 \le j \le N$, for all
	$1 \le d \le E^e_i[j]$, $\mathtt{Retrieve}$ of $\text{VID}^d_j$ finishes.
	By Lemma \ref{thm:correct-estimation}, there exists $q \in H$ such that
	$V^e_q[j] \ge E^e_i[j]$, and the $q$-th node (which is correct) reports that
	$\text{VID}^t_j$ has $\mathtt{Complete}$d for all $1 \le t \le V^e_q[j]$.
	By VID Availability property (Theorem \ref{thm:avid-availability}), the
	$\mathtt{Retrieve}$s will eventually finish, so this step will finish.

	\textit{Phase 2, steps 5.} This is local computation and will finish instantly.
\end{proof}

\begin{theorem}[Validity]
	\label{thm:validity} All blocks proposed by correct nodes are eventually delivered by
	all correct nodes.
\end{theorem}
\begin{proof}
	Assume the $i$-th node (which is correct) proposes block $B^e_i$ in epoch $e$.
	The $i$-th node invokes $\mathtt{Disperse}(B^e_i)$ on $\text{VID}^e_i$. By VID
	Termination property (Theorem \ref{thm:avid-termination}), eventually all correct
	nodes will see $\text{VID}^e_i$ $\mathtt{Complete}$. So there must exist
	an epoch $t$ where for all $j \in H$, $V^t_j[i] \ge e$. That is, in epoch $t$,
	all correct nodes report that the $i$-th node has at least dispersed into $\text{VID}^1_i$
	to $\text{VID}^e_i$. By Lemma \ref{thm:correct-estimation}, for all $j \in H$,
	$E^t_j[i] \ge e$. According to phase 2, steps 4--5, all correct nodes either have
	already delivered $B^e_i$ in previous epochs, or will deliver $B^e_i$ in epoch $t$.
\end{proof}

\begin{theorem}[Agreement and Total Order]
	\label{thm:agreement} Two correct nodes deliver the same sequence of blocks. 
\end{theorem}
\begin{proof}
	Let $i, j \in H$.
	We prove this theorem by induction on the number of epochs
	the $i$-th and the $j$-th nodes have finished. In other words,
	we prove that for any $t \ge 0$, the $i$-th and the $j$-th nodes deliver the same sequence
	of blocks in the first $t$ epochs.

	\textit{Initial ($t=0$).} Both nodes have not delivered any block. So the hypothesis clearly holds
	in this situation.

	\textit{Induction step.} Assume our hypothesis holds for $t = e-1$ ($e \ge 1$). We now prove
	the hypothesis holds for $t=e$. We first show the two nodes commit the same sequence
	of blocks with BA. By Lemma \ref{thm:counter-agreement}, $S^e_i = S^e_j$ and
	$E^e_i = E^e_j$. According to phase 2, step 1, both nodes will invoke $\mathtt{Retrieve}$
	on the same set of VIDs. By VID Correctness property (Theorem \ref{thm:avid-correctness}),
	they will get the same set of blocks and deliver them in the same order in phase 2, step 2.

	We now show the two nodes commit the same sequence of blocks with inter-node linking.
	The local variable $\mathtt{Delivered}$ stores whether a node has delivered
	a block (phase 2, steps 2, 4). By the induction hypothesis,
	the two nodes have delivered the same sequence of blocks prior to epoch $e$, so
	the variable $\mathtt{Delivered}$ is the same on the two nodes. By Lemma
	\ref{thm:counter-agreement}, $E^e_i = E^e_j$. So the two nodes will invoke
	$\mathtt{Retrieve}$ on the same set of VIDs in phase 2, step 4 and get the same set
	of blocks. Both nodes sort the blocks deterministically and deliver them in the same
	order in phase 2, step 5.

	We have proved that the $i$-th and the $j$-th nodes deliver the same sequence of blocks
	in epoch $e$. By our induction hypothesis, they deliver the same sequence until epoch $e-1$.
	So they deliver the same sequence in the first $e$ epochs. This completes the induction.
\end{proof}

\end{document}